\documentclass[preprint2]{aastex}

\usepackage{graphicx,epstopdf}

\usepackage{color}

\def\BB{{\bf {B}}}

\def\ee{{\bf {e}}}

\shorttitle{The Sun's open-closed flux boundary}
\shortauthors{Pontin \& Wyper}

\begin{document}

\title{The effect of reconnection on the structure of the Sun's open-closed flux boundary}



\author{D.~I.~Pontin}
\affil{Division of Mathematics, University of Dundee, Dundee, UK}
\email{dpontin@maths.dundee.ac.uk}

\author{P.~F.~Wyper} 
\affil{Heliophysics Science Division, NASA Goddard Space Flight Center, 8800 Greenbelt Rd, Greenbelt, MD 20771}
\email{peter.f.wyper@nasa.gov}

\begin{abstract}
Global magnetic field extrapolations are now revealing the huge complexity of the Sun's corona, and in particular the structure of the boundary between open and closed magnetic flux. Moreover, recent developments indicate that magnetic reconnection in the corona likely occurs in highly fragmented current layers, and that this typically leads to a dramatic increase in the topological complexity beyond that of the equilibrium field. In this paper we {use static models to} investigate the consequences of reconnection at the open-closed flux boundary (``interchange reconnection") in a fragmented current layer. We demonstrate that it leads to efficient mixing of magnetic flux (and therefore plasma) from open and closed field regions. This corresponds to an increase in the length and complexity of the open-closed boundary. Thus, whenever reconnection occurs at a null point or separator of this open-closed boundary, the associated separatrix arc of the so-called {\it S-web} in the high corona becomes not a single line but a band of finite thickness within which the open-closed boundary is highly structured.  This has significant implications for the acceleration of the slow solar wind, for which the interaction of open and closed field is thought to be important, and may also explain the coronal origins of certain solar energetic particles. The topological structures examined contain magnetic null points, separatrices and separators, and include a model for a pseudo-streamer. The potential for understanding both the large scale morphology and fine structure observed in flare ribbons associated with coronal nulls is also discussed.
~
\end{abstract}



\section{Introduction}

It is well established that the magnetic field that permeates the solar corona has a highly complex structure. Although it is very difficult to measure directly the magnetic field vector in the corona, this complexity can be inferred from observations of the line-of-sight magnetic field at the photosphere. With each new satellite mission that is launched, we observe photospheric magnetic flux concentrations on ever smaller scales \citep[that seem to exhibit a power-law distribution with size,][]{parnell2009}. Magnetic field extrapolations based on these observed photospheric polarity distributions exhibit an often bewildering degree of complexity. Understanding the evolution of such a complex magnetic field structure is a major challenge. 


In recent years, significant progress has been made in developing tools {with} which to characterise the coronal magnetic field. One approach involves segregating the photospheric magnetic field into discrete flux patches. This then allows the corona to be divided into distinct domains, each defined by the flux connecting pairs of these patches. Between these coronal flux domains are \emph{separatrix surfaces}, that emanate from magnetic null points. The intersection of two such separatrix surfaces forms a \emph{separator} field line -- a field line that connects two null points and lies at the intersection of four flux domains. Indeed, magnetic field extrapolations {reveal} the presence of a web of null points, separatrix surfaces, and separators that form a \emph{skeleton}  based upon which the magnetic connectivity of {the coronal field} may be understood \citep[e.g.][]{longcope2005b,regnier2008}. The separatrix surfaces of this skeleton represent locations at which the mapping between boundary points via the magnetic field lines exhibits discontinuities. Also of interest are layers in which this field line mapping exhibits strong (but finite) gradients. These are known as \emph{quasi-separatrix layers} (QSLs), being regions at which the \emph{squashing factor}, $Q$ \citep{titov2002}, is large.

{Null points, separators, and QSLs}, at which the field line mapping is either discontinuous or varies rapidly, are of interest not only in analysing the structure of the coronal magnetic field, but for understanding its dynamics. This is because these locations are prime sites for the formation of current layers at which magnetic reconnection may occur, releasing stored magnetic energy \citep[][and references therein]{pontin2012a}. In particular, they have been implicated in the formation of current sheets associated with solar flares, jets, and coronal mass ejections \citep[e.g.][]{fletcher2001,mandrini2006,barnes2007,titov2008,lynch2008,pariat2010,sun2013}. One major piece of supporting evidence is the coincidence of flare ribbons with footpoints of separatrix and QSL field lines in coronal field extrapolations \citep{masson2009,janvier2014}.

One particular location at which the magnetic field line mapping is discontinuous is at the interface between closed and open magnetic flux, i.e.~the boundary between magnetic field lines that are anchored at both ends at the photosphere, and those that extend out into the heliosphere. Magnetic reconnection at this open-closed flux boundary is one of the principal mechanisms proposed to explain the properties of the slow, non-steady, solar wind \citep[e.g.][]{fisk1998}. The slow solar wind is characterised by strong fluctuations in both velocity and plasma composition, the latter of which is consistent with the wind being composed of some component of closed-flux coronal plasma \citep[e.g.][]{hansteen2012}. Reconnection specifically at the open-closed flux boundary is also implicated in the generation of impulsive \emph{solar energetic particle (SEP)} events, due to the observed ion abundances of these events \citep{reames2013}.

Typically, computational models of the Sun's global magnetic field exclude the outflowing plasma of the solar wind, but include its effect  by imposing a magnetic field that  is purely radial at some height above the photosphere (termed the `source surface'). Excluding all contributions to the solar magnetic field other than the global dipole, the coronal field is characterised by two polar coronal holes of open magnetic field lines and a band of closed flux around the equator, these two being separated by separatrix surfaces that meet the base of the heliospheric current sheet (HCS) at the source surface. The question arises: when the full complexity of the coronal field is introduced, what is the nature of the boundary between open and closed flux? {In a series of papers, Fisk and co-workers \citep[e.g.][]{fisk1998,fisk2001,fisk2005} developed a model for the dynamics of the Sun's open magnetic flux, that was also used to explain the acceleration of the solar wind mediated by reconnection between open and closed field lines \cite[termed {\it interchange reconnection} by ][]{crooker2002}. In their model, open field lines can freely mix with and diffuse through the closed field regions, and indeed it is predicted that this open flux component should become uniformly distributed throughout the (predominantly) closed field region \citep{fisk2006}. While noting that such a scenario requires the presence of current sheets in the corona between open and closed flux, these studies do not address the magnetic field structure in detail.
Indeed,} the topological admissibility of such free mixing of open and closed flux has since been questioned \citep{antiochos2007}, making it difficult to reconcile the interchange reconnection {solar wind acceleration} mechanism with the broad observed latitudinal extension of the slow solar wind streams (up to $60^\circ$, especially at solar minimum) \citep[e.g.][]{mccomas2000}. Nonetheless, recent modeling of the global coronal magnetic field has suggested a resolution to the apparent contradiction that plasma that appears to originate in the closed corona is observed far from the HCS at large radii. It has been demonstrated that additional regions of open flux that are disconnected from the polar coronal holes (at the photosphere) may {indeed} exist. The distinct photospheric regions of open magnetic flux are partitioned by multi-separatrix structures associated with multiple nulls points, typically comprising a dome-shaped separatrix enclosing the closed flux between the two open field regions, intersecting with a vertical  \emph{separatrix curtain} \citep{titov2011,platten2014}.  Even when coronal holes are not disconnected, there may exist very narrow channels of open magnetic flux at the photosphere connecting two larger open flux regions. In this case the narrow channel is associated with a QSL curtain. Both the QSL and separatrix curtains extend out into the heliosphere, and have been shown to  map out a broad latitudinal band around the HCS, termed the \emph{S-web} \citep{antiochos2011,crooker2012}. The corresponding arc structures at the source surface in global models are associated with pseudo-streamers in the observations, and there is growing evidence that these structures are associated with slow solar wind outflow \citep[e.g.][]{owens2013}.

The above studies have revealed that the open-closed flux boundary has a complex topological structure involving null points and their associated separatrices, separators and QSLs. Moreover, it has been recently demonstrated that when reconnection occurs in astrophysical plasmas, the 3D topological complexity can dramatically increase beyond that of the equilibrium field \citep[e.g.][]{daughton2011,wyper2014a,wyper2014b}. In this paper we use simple {static magnetic field} models to investigate the implications of reconnection for the magnetic field connectivity at the open-closed flux boundary when the reconnecting current layer exhibits a fragmented structure, expected to be the typical case in the corona. The paper is structured as follows. In Section \ref{3dtearsec} we summarise recent relevant results on current layer instabilities. In Sections \ref{domesec} and \ref{curtainsec} we investigate the topological effect of reconnection in configurations defined by an isolated separatrix dome and a separatrix curtain.
In Sections \ref{discusssec} and \ref{concsec}, respectively, we present a discussion of the results and our conclusions.

\section{Non-linear tearing during 3D reconnection}\label{3dtearsec}
In recent years a major advance in our understanding of magnetic reconnection has been the realisation that the reconnection rate can be substantially enhanced when the current layer breaks up in response to a tearing instability. While the linear phase of the classical tearing mode is slow, non-linear tearing in two dimensions (2D) via the \emph{plasmoid instability} can grow explosively, and  lead to a reconnection rate that is only weakly dependent on the resistivity \citep{loureiro2007,bhattacharjee2009,loureiro2013}. In 2D as the instability proceeds a myriad of magnetic islands are formed as current layers fragment into chains of X-type and O-type nulls. The conditions for onset of the instability are that the (inflow) Lundquist number be $\gtrsim10^4$, and the current layer  aspect ratio be $\gtrsim 50$.

\begin{figure}[t]
\centering
\includegraphics[width=6cm]{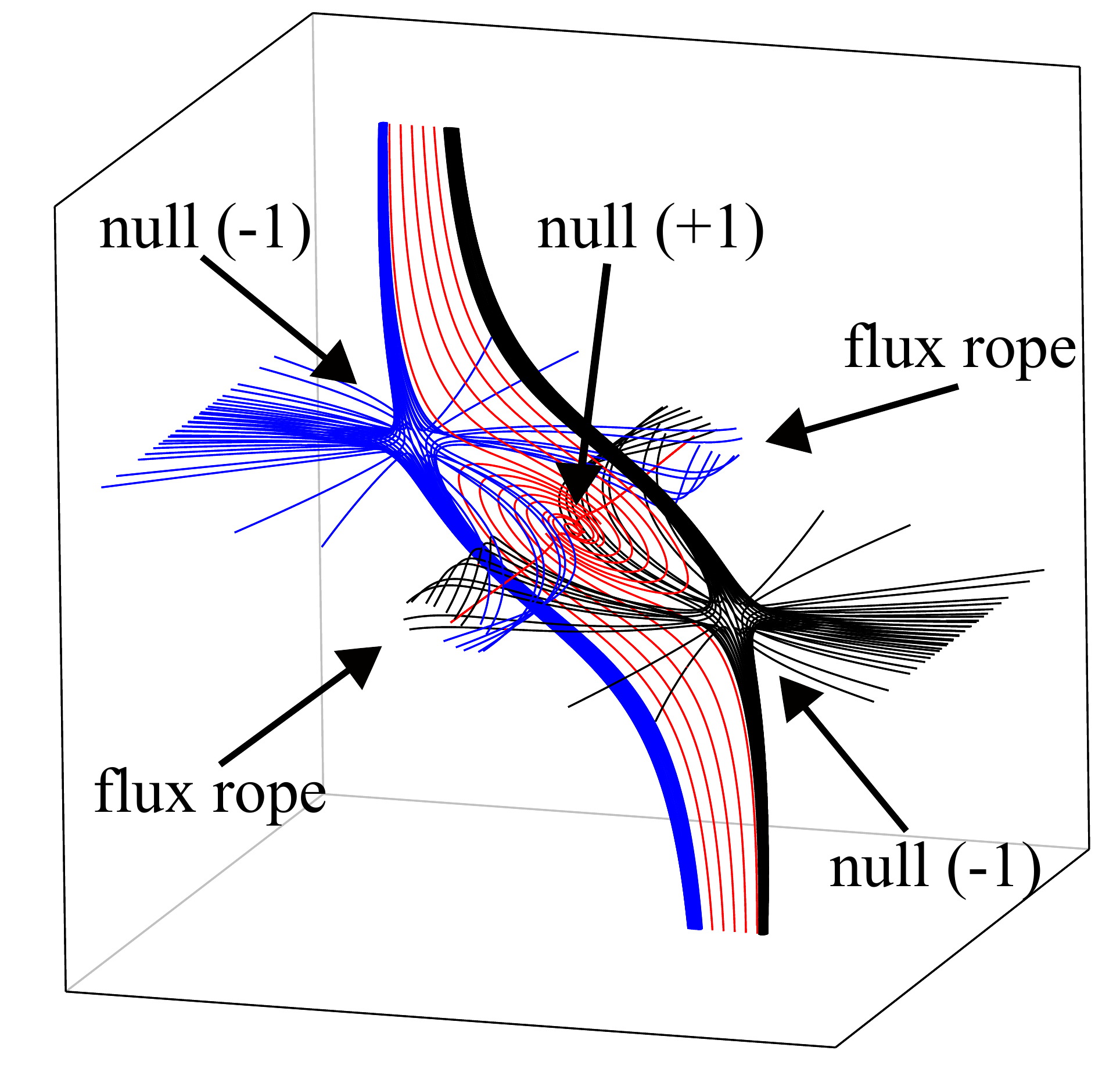}
\caption{A model magnetic field showing the magnetic topology following the initial bifurcation of the central 3D null following the tearing instability -- after \cite{wyper2014a}.}
\label{WPtop}
\end{figure}
It has recently been demonstrated that the plasmoid instability also occurs in three-dimensional (3D) current layers. \cite{daughton2011} performed 3D particle-in-cell simulations of an initially planar, infinite current layer. They noted that magnetic `flux ropes' formed in place of the magnetic islands from the 2D picture with flux often threading in and out of multiple flux ropes. 
\cite{wyper2014a,wyper2014b} by contrast studied MHD simulations of a 3D magnetic null point undergoing external shear driving. They observed the initial formation of a laminar current layer centred on the null, which was found to become unstable at a threshold similar to the 2D case (Lundquist number $\gtrsim 2\times 10^4$, aspect ratio $\gtrsim 100$). This threshold onset condition is very likely to be exceeded for typical current sheets formed in the corona. In this study, it was demonstrated that the onset of tearing leads to the creation of new 3D null points in bifurcation processes. In particular, 3D spiral nulls are formed that are the analogue of 2D islands, and the spine lines of each of these nulls forms the axis of a pair of magnetic flux ropes, as shown in Figure \ref{WPtop}. Crucially, and in contrast to the 2D case, these flux ropes are open structures -- they are not surrounded by flux surfaces as 2D islands are (this is required on topological grounds due to the variation of $\BB$ along the direction of the flux rope axis and the solenoidal condition on $\BB$). As a result, no new isolated domains of magnetic connectivity are formed. Rather, the flux ropes are composed of a mixture of flux from the two connectivity domains (flux located above and below the separatrix of the single null prior to the instability), wrapped around each other. The result is that, when the field line connectivity is analysed, a layer is found  in which magnetic flux from the two connectivity domains is rather efficiently mixed -- in spiral patterns associated with the multiple flux rope structures. Our intention here is to investigate the effect on the global connectivity when the fragmented, reconnecting current layer is embedded in some generic coronal field structures.

\section{Isolated dome topology}\label{domesec}
\begin{figure}
\centering
(a)\includegraphics[width=7cm]{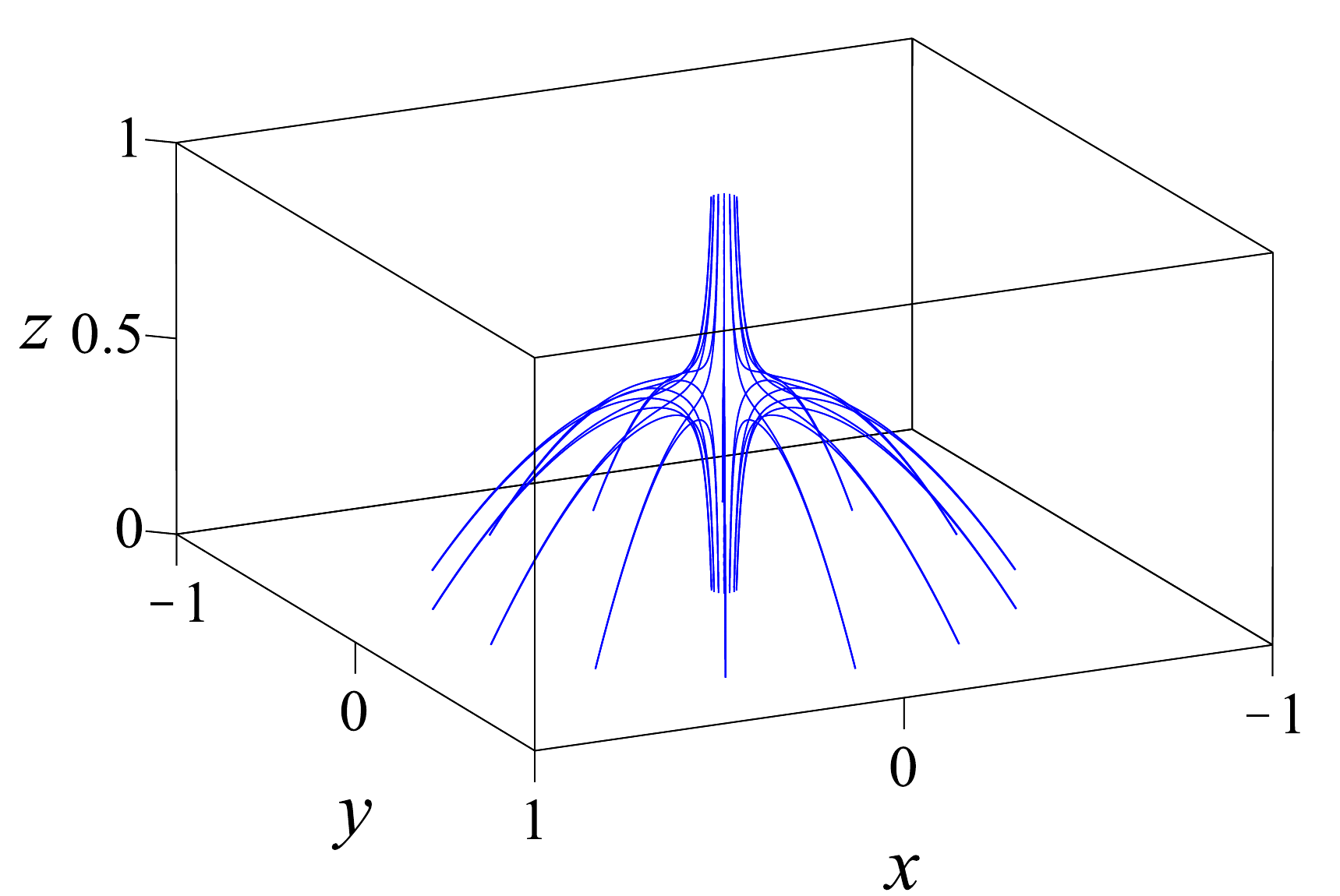}
(b)\includegraphics[width=7cm]{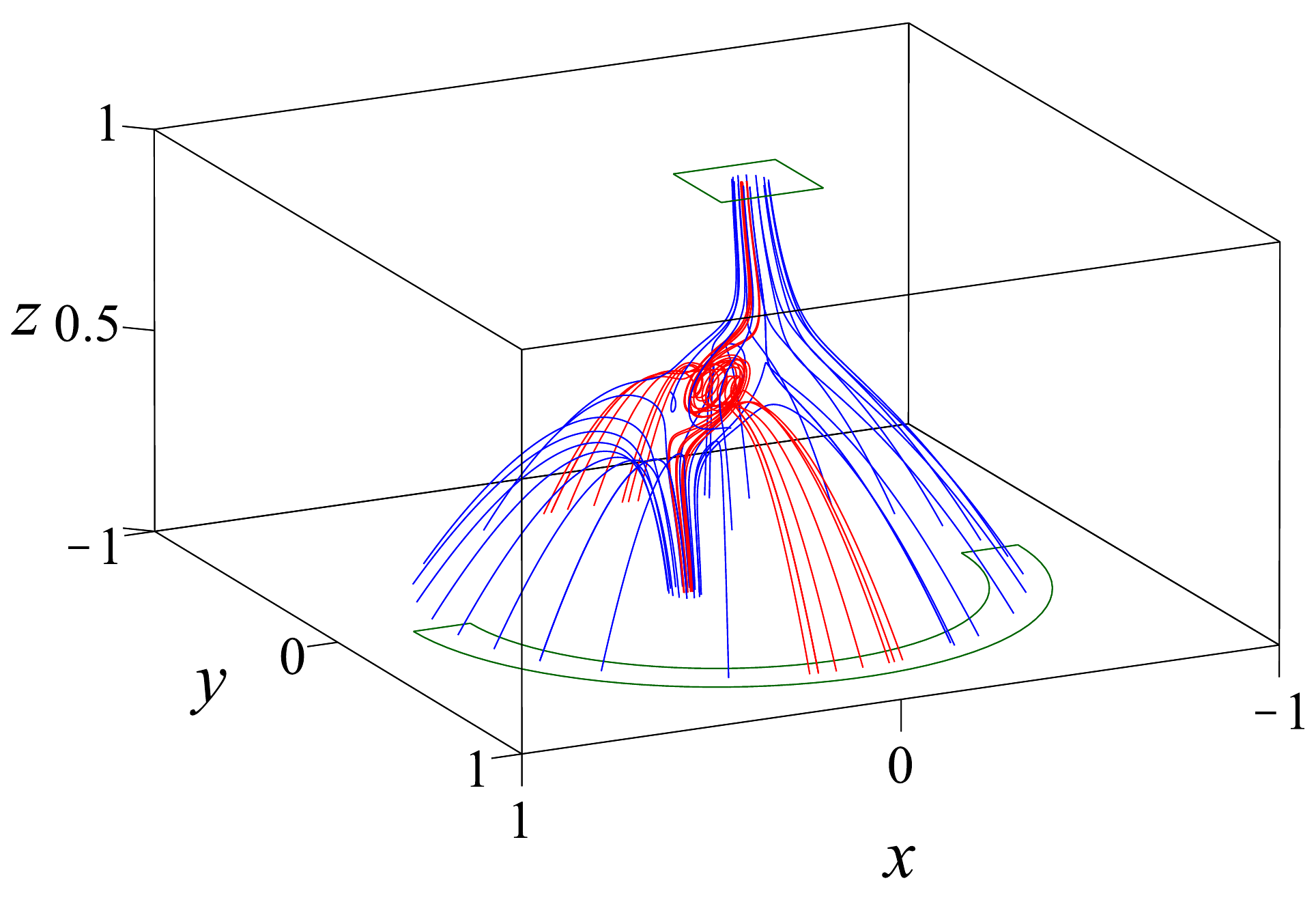}
\caption{(a) Magnetic field lines for the isolated dome topology given by Equation (\ref{bdome}), and (b) after the addition of one flux ring (state 1a) -- sample field lines in the flux ropes are coloured red. {Green boxes outline the regions in which connectivity and $Q$ maps are calculated in Figures \ref{dome_oc_map}, \ref{dome_q_map}.}}
\label{dometop}
\end{figure}
We first examine the simplest generic coronal configuration containing a 3D null point -- the isolated dome topology (see Figure \ref{dometop}(a)). Such a configuration is always present, for example, when a photospheric region of one magnetic polarity is embedded in a region of opposite polarity. {Such 3D nulls are preferred sites of current sheet formation \citep{pontincraig2005,pontinbhat2007a}}, and the significance of reconnection in an isolated dome configuration has been considered, for example, by \cite{antiochos1996,pariat2009,edmondson2010}. As demonstrated by \cite{pontin2013}, spine-fan reconnection in such a dome topology is characterised by a transfer of flux in one side of the separatrix dome and out the other side, which can be driven dynamically or occur during a relaxation process as the coronal field seeks a minimum energy state.

In order to examine the effect of tearing of the null point current sheet on the field structure we consider the following simple model. The fields that we construct are not equilibrium fields (e.g.~are not {force-free}) -- however this is of no importance since we do not consider here any dynamical processes. Rather, our purpose is solely to examine the field topology/geometry that results from breakup of a reconnecting current layer. We consider the (dimensionless) field 
\begin{equation}\label{bdome}
\BB=x\ee_x+y\ee_y+\left(1-2z-4r^2\tanh^2(4r)\right)\ee_z,
\end{equation}
where $r=\sqrt{x^2+y^2}$. This magnetic field contains a null point at $(x,y,z)=(0,0,0.5)$ above a photospheric plane represented by $z=0$, see Figure \ref{dometop}(a). {The isolated dome topology appears over a wide range of scales in the corona. Hereafter we discuss any length scales in terms of a characteristic `macroscopic' length scale of the overall structure, that we denote $D$. In our model field both the separatrix dome diameter at $z=0$ and the null point height are of order 1, so for the model field $D\sim 1$. From magnetic field extrapolations it is observed that in the corona this scale $D$ of the dome separatrix structure can be as large as a few hundred Mm \citep[usually in the vicinity of active regions, e.g.][]{delzanna2011,platten2014}, and at least as small as a few tens of km \citep[in quiet-Sun regions, where the lowest null point height in extraploations is likely limited by the magnetogram resolution, e.g.][]{regnier2008}.}

Onto the `background' field (\ref{bdome}) we super-impose a magnetic flux ring to simulate the topological effect of magnetic reconnection occurring at some particular location in the volume. This method is motivated and described in detail in \cite{wilmotsmith2011a,pontin2013}. The field of this flux ring is taken of the form
\begin{eqnarray}\label{b_ring}
\BB_R\!&=&\!\! B_0\,\nabla\times\left(  A_y {\bf e}_y\right),\\
A_y \!&=&\!\! \exp\left(-\frac{(x-x_N)^2}{L^2} - \frac{(y-y_N)^2}{l^2} - \frac{(z-z_N)^2}{L^2}\right). \nonumber
\end{eqnarray}
We begin by super-imposing a single flux ring of this form onto the field of Equation (\ref{bdome}), centred at the null (i.e.~we set $x_N=y_N=0$, $z_N=0.5$). {$L$ and $l$ are the characteristic size of the flux ring in the $xz$-plane (plane of $\BB_R$) and along $y$, respectively.} For $B_0$ small, the effect of adding the flux ring is to collapse the spine and fan of the null point towards one another, as described by \cite{pontin2013}. This has the effect of transferring flux in one side of the dome and out of the other, and is consistent with the topology of a single laminar reconnection layer at the null \citep{pontinbhat2007a,galsgaard2011a}. However, for larger values of $B_0$ the field becomes elliptic at $(x,y,z)=(x_N,y_N,z_N)$ as the flux ring field dominates over the hyperbolic background field. The result is a bifurcation of the original null into three null points as described above in Section \ref{3dtearsec} -- see Figure \ref{WPtop} -- and the generation of a pair of flux ropes (see Figure \ref{dometop}(b)). This models the magnetic topology when the current layer undergoes a spontaneous tearing instability as observed by \cite{wyper2014a}. Parameters for the magnetic field -- denoted state 1b -- are presented in Table \ref{tbl}.

\begin{figure}
\centering
(a)\includegraphics[width=7cm]{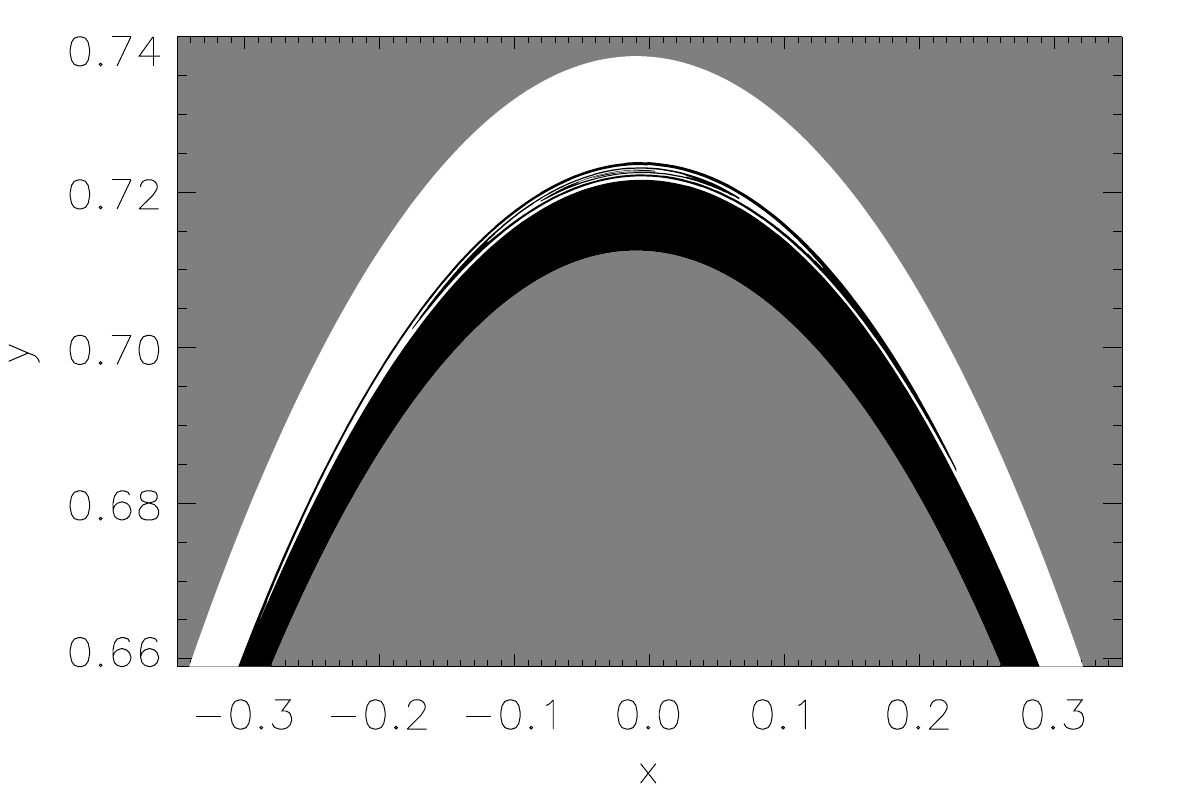}
(b)\includegraphics[width=7cm]{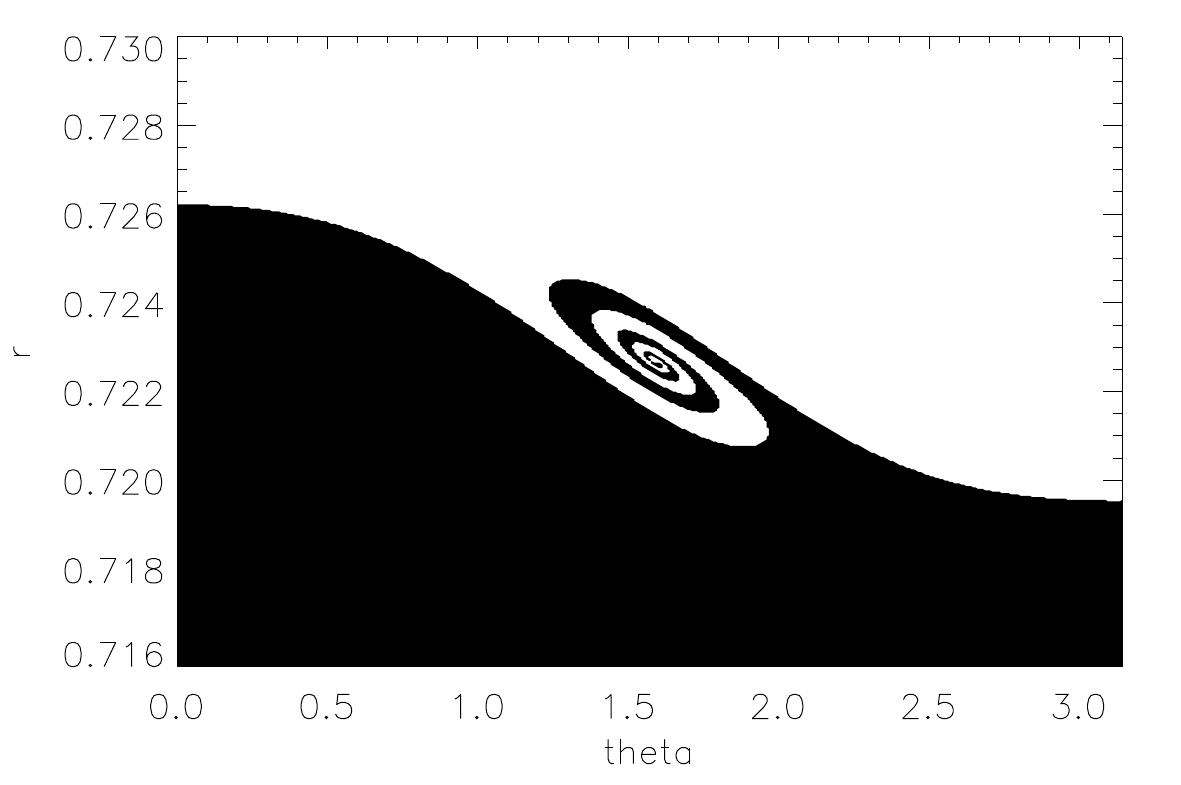}
(c)\includegraphics[width=7cm]{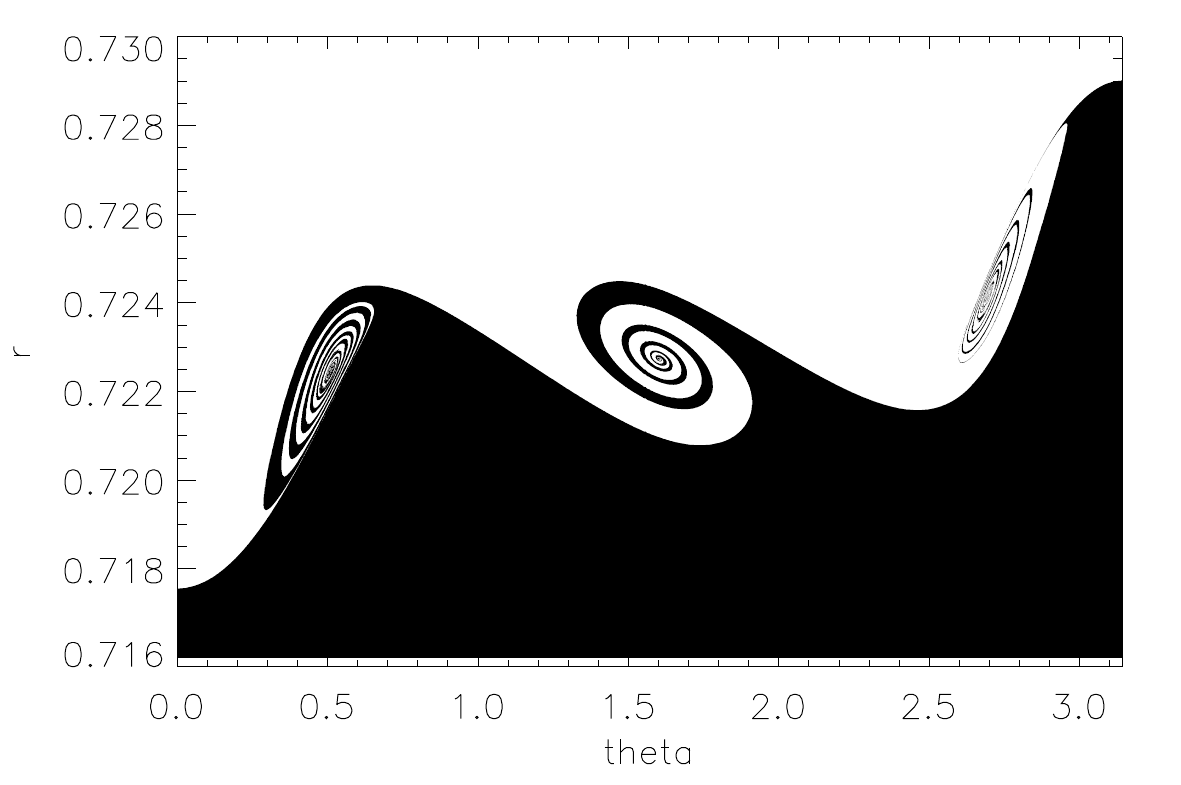}
\caption{Photospheric connectivity maps for the isolated dome topology. White regions contain footpoints of open field lines, black of closed field lines (grey: field lines from these points not traced). (a) State 1b, and (b) the same plot in standard $r-\theta$ coordinates centred at $(x,y)=(-0.01,0)$. (c) State 1c containing three flux rings.}
\label{dome_oc_map}
\end{figure}

In order to visualise the new field structure created after tearing onset we first plot a connectivity map of field lines from the lower boundary.  That is, we trace field lines from a grid of footpoints on the lower boundary and distinguish field lines that are closed (return to the lower boundary $z=0$) and open (exit through the upper boundary $z>1$). The resulting map can be seen in Figure \ref{dome_oc_map}. While the flux ropes are approximately circular near the apex of the dome, they are compressed towards the separatrix and stretched in the azimuthal direction by the global field geometry, and thus appear as flattened spiral structures in the connectivity map (Figure \ref{dome_oc_map}(a)). In order to more clearly visualise this structure we reproduce the map in a polar coordinate system in Figure \ref{dome_oc_map}(b). The observed spiral pattern of mixing of open and closed flux reproduces the behaviour in the dynamic MHD simulations -- see Figure 8 of \cite{wyper2014a}.
One can include the effect of a further breakup of the current layer through the inclusion of additional flux rings. Adding two such flux rings centred at the non-spiral nulls of state 1a leads each of these nulls to undergo a bifurcation, resulting in a total of seven nulls. This naturally introduces additional spiral structures in the field line mapping, as show in Figure \ref{dome_oc_map}(c) (state 1c, for parameters see Table \ref{tbl}), and if one were to iterate this procedure by adding more flux ropes a mapping with complexity of the order of that seen in the MHD simulations of \cite{wyper2014a} could be obtained.

\begin{figure}
\centering
(a)\includegraphics[width=7cm]{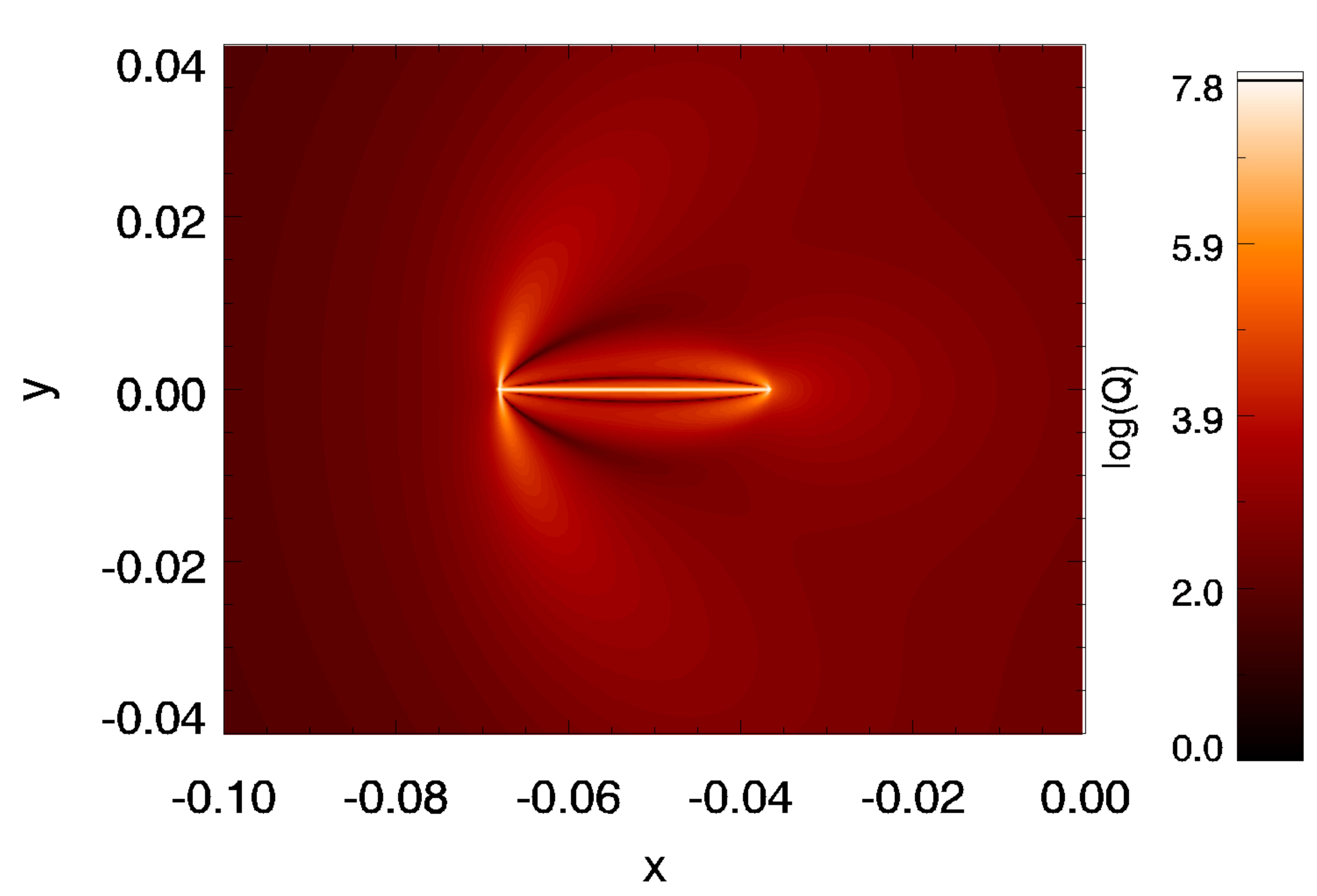}
(b)\includegraphics[width=7cm]{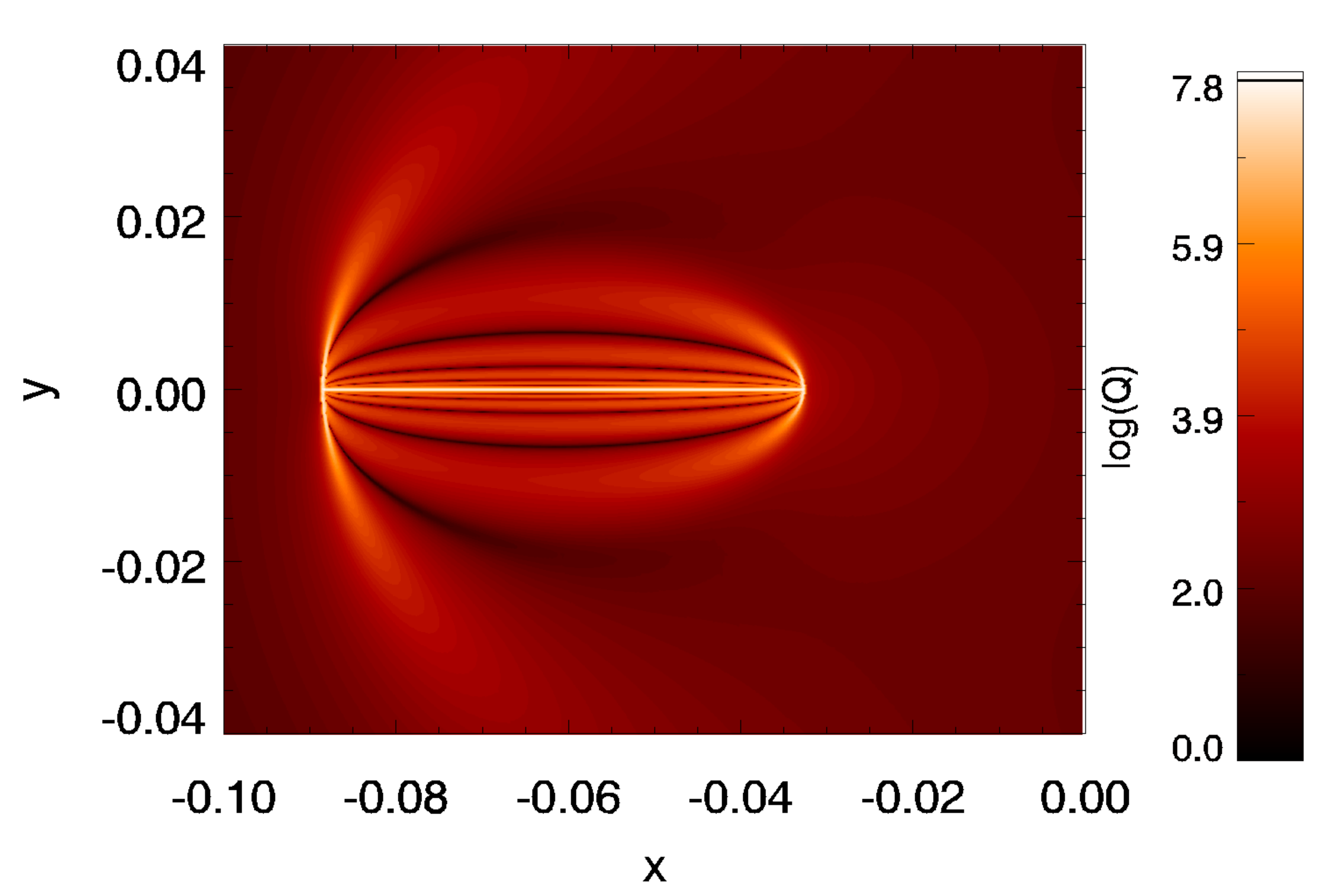}
(c)\includegraphics[width=7cm]{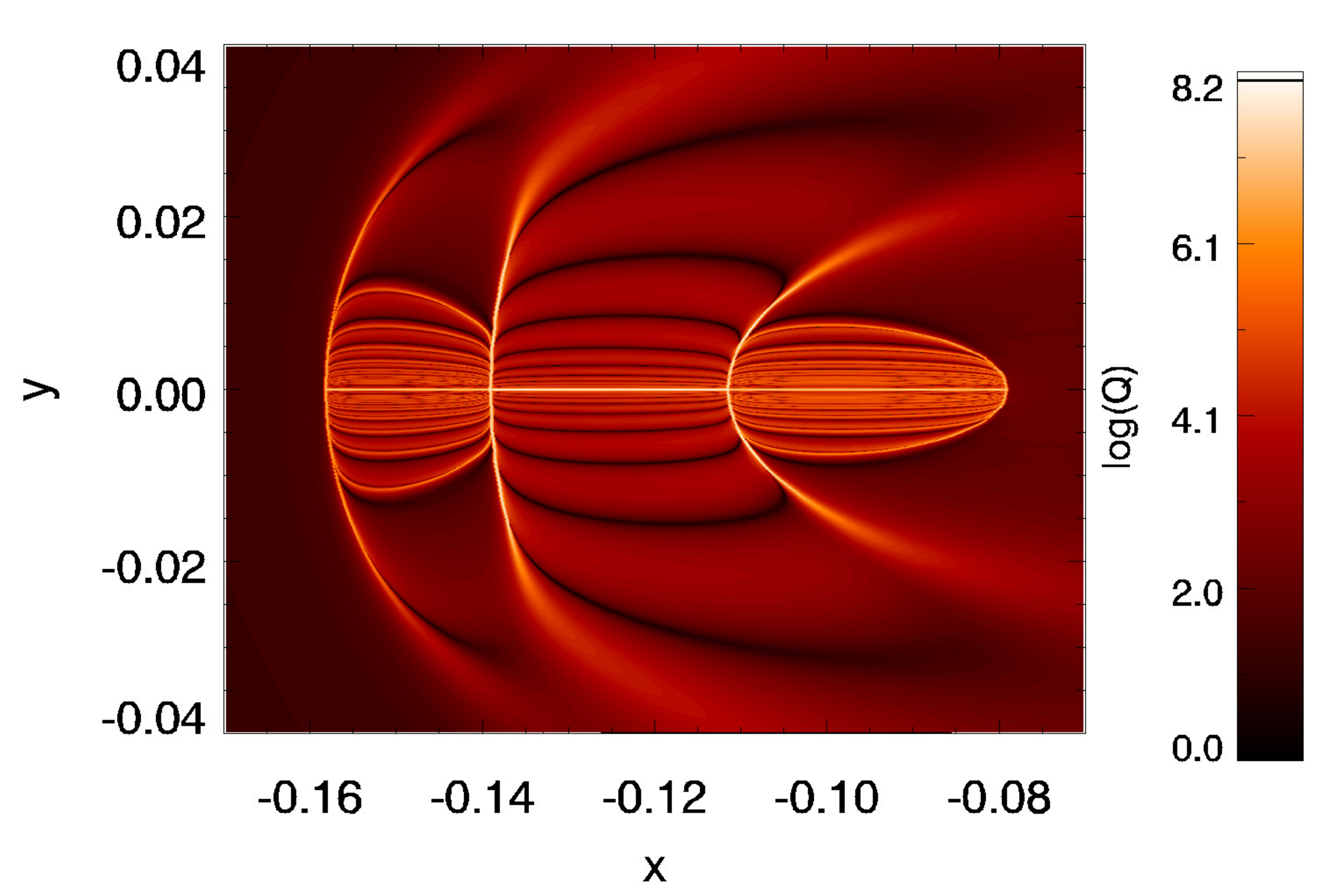}
\caption{Plot of $Q$ on the top boundary $z=1$, for (a) state 1a, (b) state 1b, and (c) state 1c.}
\label{dome_q_map}
\end{figure}

We now turn to consider the characteristics of the open flux that exits the domain through the top boundary at $z>1$. Since all of this flux is open, a connectivity map does not reveal this structure. However, we can use for example the \emph{squashing factor}, $Q$, to visualise the field line mapping from $z=0$ to $z=1$. Here we plot $Q$ on the surface $z=1$. The distribution of $Q$ is obtained by integrating field lines from a rectangular grid of {typically around $10^6$} footpoints and then calculating the required derivatives using finite differences over this grid. $Q$ is formally infinite on spine and fan field lines since they represent discontinuities in the field line mapping. However, calculating $Q$ numerically as we do here they show up only as sharp points and lines, respectively, with very high values of $Q$. One should therefore not attach physical meaning to the maximum value of $Q$ in the plots (attained at the separatrix/spine footpoints) as it is determined entirely by the resolution of the field line grid.

\begin{figure*}
\centering
\includegraphics[width=0.99\textwidth]{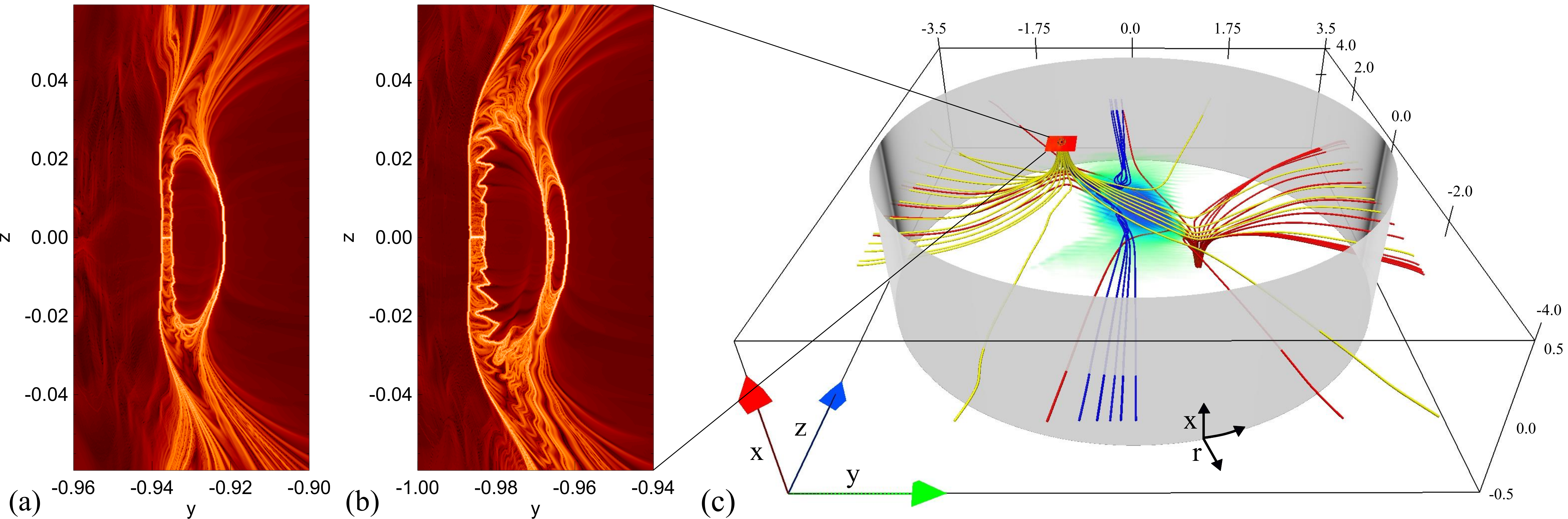}
\caption{(a,b) $Q$ on the top boundary of the simulation of \citet{wyper2014b} at two times soon after tearing occurs. (a) $t=13.8$. (b) $t=14.4$. (c) Magnetic field line structure at $t=14.4$. Blue field lines show one pair of flux ropes. The grey iso-surface shows the cylindrical surface used to calculate $Q$. Volume shading indicates the current density modulus.}
\label{sim_q_map}
\end{figure*}

In the background dome topology of Equation (\ref{bdome}) a single spine line intersects the $z=1$ boundary, and the $Q$-map displays a single maximum at the origin. When a null point bifurcation occurs during reconnection, the topological structure changes to that shown in Figure \ref{WPtop}. In this case a vertical separatrix extends up to the top boundary, bounded on either side by a pair of spine lines. Examining the $Q$ distribution on $z=1$ for state 1a (Figure \ref{dome_q_map}a), the separatrix footprint is clearly in evidence {(horizontal line of high $Q$)}. Increasing the strength of the flux ring in the model (state 1b) leads to a lengthening of this separatrix due to the increased separation of the nulls (Figure \ref{dome_q_map}b). A further breakup of the current sheet leads to the appearance of multiple vertical separatrices, as {seen} in Figure \ref{dome_q_map}(c).

There are two additional noteworthy features of the $Q$ distributions. First, note the arcs of high 
and low $Q$ that run parallel to the separatrix footprint. These become more pronounced and numerous as the flux ring strength is increased (compare Figures \ref{dome_q_map}(a,b)). Their origin can be understood as follows. Consider field lines traced down from the top boundary that enter one of the flux ropes. Some local bundles of field lines will spiral around the rope axis and then `leave' the flux rope at its top or bottom (in $z$) with a range of values of $\theta$ but roughly constant $z$ values ($\theta$ being the azimuthal angle in the $xy$-plane). Due to their range of $\theta$ values on leaving the rope they diverge in the azimuthal direction as they are traced onwards to the lower boundary -- and therefore exhibit relatively high $Q$ values. By contrast, adjacent field lines that leave the flux rope along its sides at approximately equal $\theta$ values but differing $z$ values are naturally squeezed in towards the fan as they approach the photosphere -- they do not diverge with the null point fan geometry owing to their close alignment in the $\theta$ direction. This leads to a lower $Q$. For a stronger flux ring (more substantial flux rope) field lines have the opportunity to spiral multiple times around the rope axis, leading to multiple $Q$ stripes. The second feature to note are the high-$Q$ ridges emanating from each spine footpoint in the $y$-direction, that are present for the following reason. Adding the flux rings naturally generates a strong field component in the $y=0$ plane. Therefore the two non-spiral nulls have quite asymmetric fan eigenvalues (their ratio is around 2.5 in state 1a). The weak field direction corresponds to the $y$-direction, and it is natural that $Q$ is largest in this weak-field region of diverging fan field lines.


To test the robustness of the structures in $Q$ described above, two $Q$-maps were calculated using magnetic fields taken from the {dynamic} MHD simulation of \citet{wyper2014b}. To avoid discontinuities in $Q$ brought on by the corners of the domain we calculated $Q$ using the foot points of field lines traced from a fixed grid on the top boundary to a cylindrical surface defined by $r = \sqrt{y^2 + z^2}=3.4$, red and grey surfaces in Fig. \ref{sim_q_map}(c) respectively.
Fig.~\ref{sim_q_map}(a) shows $Q$ on the top ``open'' boundary soon after tearing has occurred in the layer -- see also Fig.~5 of \citet{wyper2014b}. Note that $x$ is the vertical direction in the simulation domain and that the spine-fan collapse occurs in the $z=0$ plane, Fig \ref{sim_q_map}(c). At this time there is a single pair of flux ropes within the current layer. Despite significant fine structure {(likely resulting from turbulent dynamics in the outflow region) the structures {in our simple model} described above are {clearly evident also in the dynamic MHD simulation}. A {short} high-$Q$ line corresponding to a vertical separatrix surface is apparent near $(y,z)\approx(-0.935,0)$, whilst a number of parallel stripes of $Q$ can be seen extending to either side of it. Additionally, two high-$Q$ ridges emanate from the ends of this separatrix surface. 
The {observed} closed loop of high $Q$ results from the {flux rope pair being located in the reconnection} outflow {having detached from the open-closed boundary}, see the discussion of \citet{wyper2014a,wyper2014b}. This splits the field lines that connect from $x=0.5$ to $y=-3.5$ into two bundles: those that connect directly from the top boundary to the side and those that loop first around the back of the flux rope pair. The foot points of the latter are found within the loop of high $Q$. Note that the field line connectivity changes continuously around the boundary of this loop, so the value of $Q$ is large but finite.
{At {the} later time two pairs of flux ropes are present in the outflow region of the current layer resulting in an additional separatrix footprint being present, Fig.~\ref{sim_q_map}(b). The gap observed between the pair of vertical separatrix footprints (in contrast to Fig.~\ref{dome_q_map}(c)) is again a result of the detachment of the nulls from the open-closed boundary.
}

\begin{figure}
\centering
\includegraphics[width=8cm]{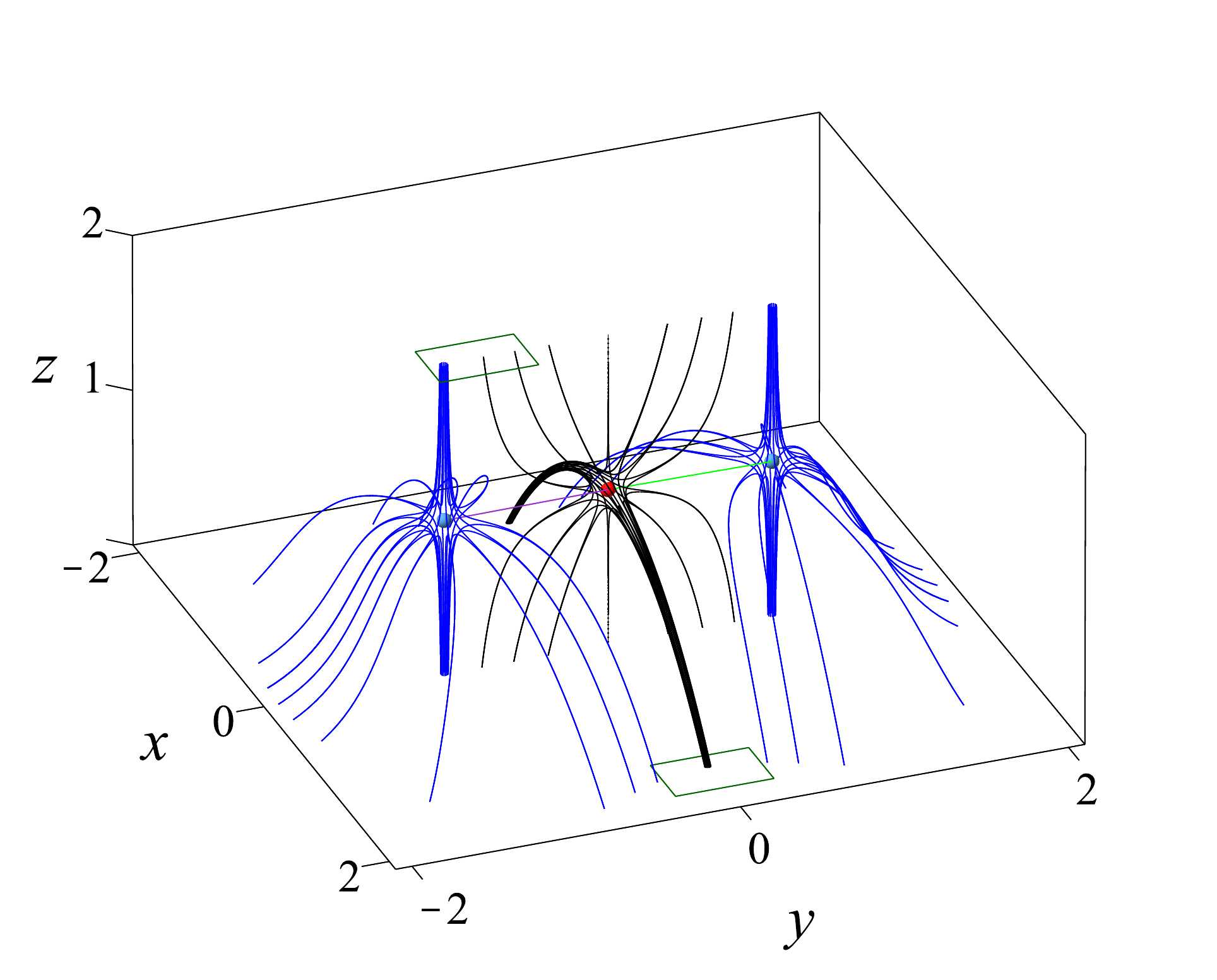}
\caption{Magnetic field lines for the separatrix curtain topology, Eq.~(\ref{bcurtain}). {Red and blue spheres correspond to nulls with topological degree of $+1$ and $-1$ respectively, whilst the two separators are shown in green and purple.} Green boxes outline the regions in which connectivity maps are calculated.}
\label{curtaintop}
\end{figure}

{We conclude that tearing of the reconnecting current layer at an isolated coronal null separatrix dome leads to the formation of an envelope around the initial dome structure in which magnetic flux from inside and outside the dome is efficiently mixed together. Additionally, vertical separatrix curtains are formed during each null point bifurcation. The implications of these results will be discussed in Section \ref{discusssec}.}

\section{Separatrix curtain / pseudo-streamer topology}\label{curtainsec}
\subsection{Magnetic field model}
Isolated separatrix dome structures associated with a single null as considered in the previous section separate small pockets of closed flux from the open flux in the polar regions {(as well as being prevalent in closed flux regions)}. However, it is also typical to have much more complicated separatix configurations separating open and closed flux. In particular, in global field extrapolations it is seen that vertical \emph{separatrix curtains}  lie between coronal holes that are {of the same polarity but are} disconnected at the photosphere \citep{titov2011,platten2014}. These curtains, together with QSLs associated with narrow corridors of open flux, are associated with arc structures at the source surface in global models that are interpreted as being associated with {\it pseudo-streamers} \citep{antiochos2011,titov2011,crooker2012}.

We consider here a simple model containing a vertical separatrix surface representing one of these curtains. This intersects a separatrix dome associated with three coronal null points {along two separator lines. The separatrix curtain consists} of the fan surface of one of these nulls (see Figure \ref{curtaintop}). 
The magnetic field expression for our model is as follows 
\begin{eqnarray}
\BB\!\!\!&=&\!\!\!
-2x\,\ee_x - y(y^2-1)\,\ee_y\nonumber\\
&&\!\!\!+\left\{(z-1)(1+3y^2)+2x^2\right.\nonumber\\
&&\!\!\!\left.-2(\tanh(6y+8.4)-\tanh(6y-8.4)-2)y^2\right\}\ee_z.\nonumber\\
&& \label{bcurtain}
\end{eqnarray}
{Again a characteristic length scale of the overall structure is of order 1 in the model field (say the null point height or separation -- see Figure \ref{curtaintop}), that we refer to as $D$. {On the Sun, $D$ is observed to be as large as {a quarter of the solar radius \citep[$\sim 170$Mm, e.g.][]{Wang2007}}, and may be at least as small as tens of kilometers in quiet Sun regions as discussed before.}}
Note that to find the separators in these models the ``progressive interpolation'' method \citep{haynes2010,close2004} was used, whereby field lines were traced from a ring encircling one of the associated null points on its fan plane to identify the approximate position of each separator, before using an iterative bisection procedure to find each separator to a desired accuracy.

\begin{figure*}[!t]
\centering
\includegraphics[width=0.8\textwidth]{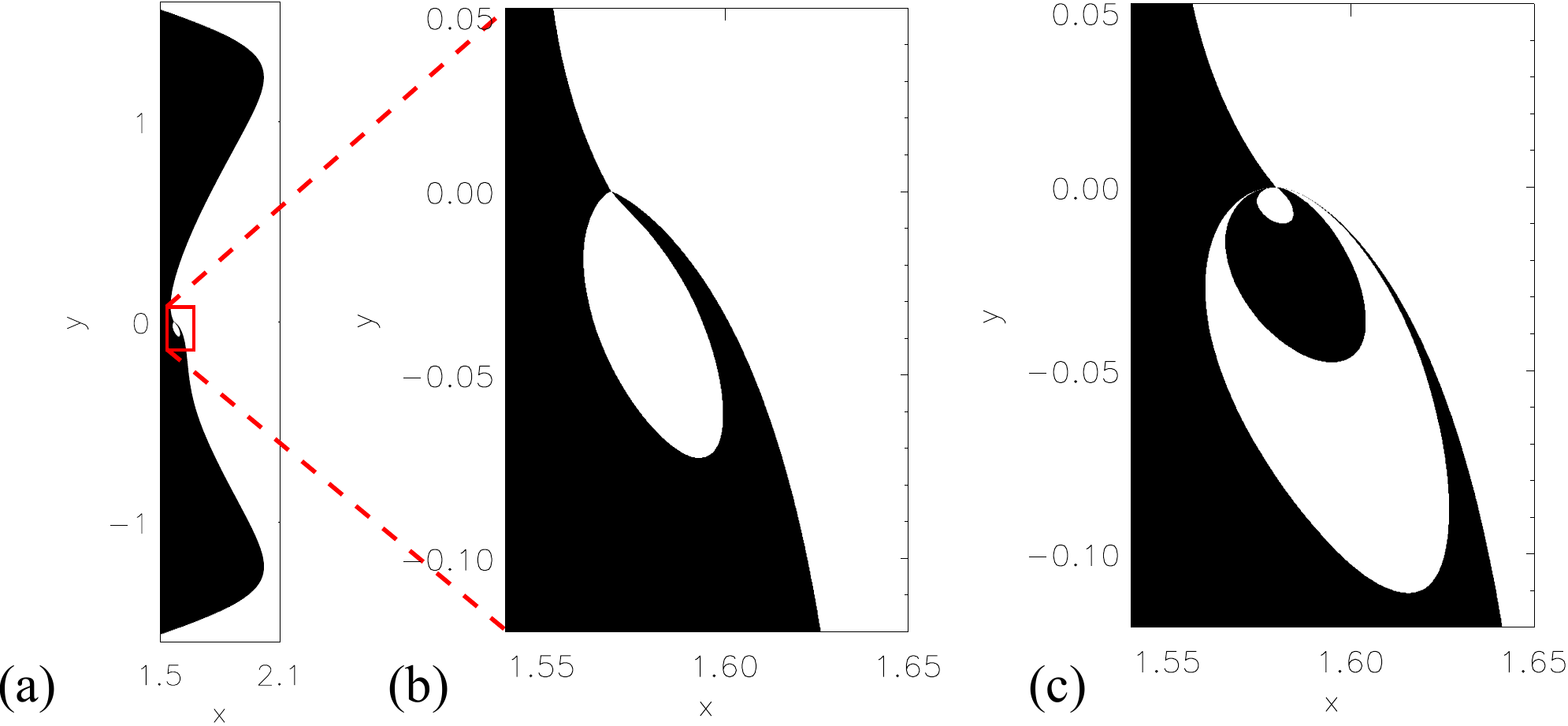}
\caption{Photospheric connectivity maps for the separatrix curtain topology. White regions contain footpoints of open field lines, black of closed field lines. (a) State 2a, (b) state 2a, close-up, (c) state 2b.}
\label{sep_oc_1rope_phot}
\end{figure*}

\begin{figure*}
\centering
\includegraphics[width=0.9\textwidth]{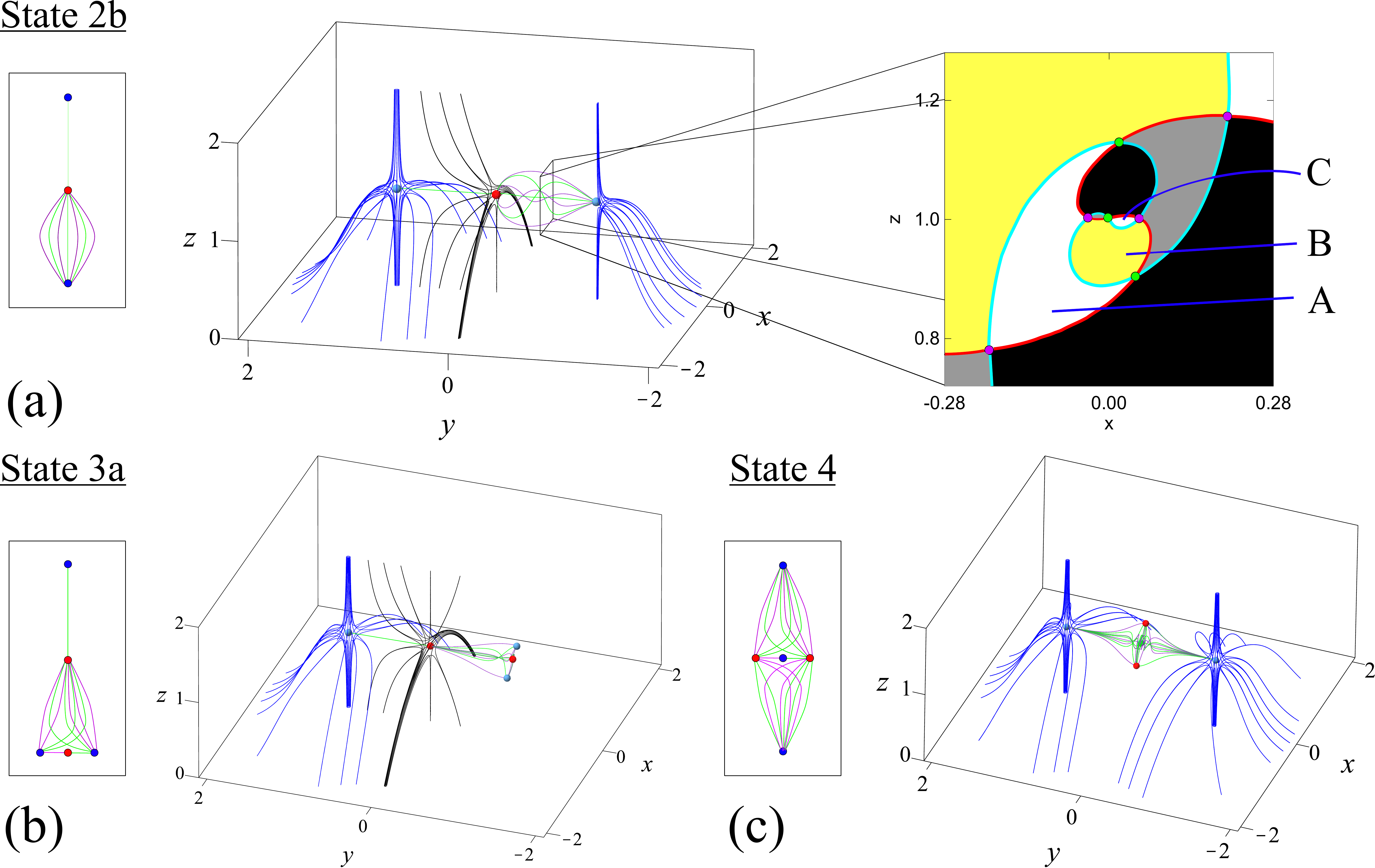}
\caption{(a) State 2b. Middle panel shows the 3D structure of the magnetic field. Red and blue spheres correspond to nulls with topological degree of $+1$ and $-1$ respectively. {The separators are alternately shown in green and purple representing their pairwise formation.} Left panel: null graph showing the connectivities of the nulls and separators. Right panel: connectivity map taken from a cut {at $y=-0.5$. Yellow and white regions correspond to open field lines that connect to $x<0$ and $x>0$ {on the photosphere}, respectively. Similarly, {grey and black regions} correspond to closed field lines that connect {from the vicinity of the spine footpoint at $(x,y,z)=(0,-1,0)$ to the vicinity of the spine footpoints of the central null at $x<0$ (grey) and $x>0$ (black)}.} The {separatrix surfaces of the two nulls (red and cyan lines)} lie at the intersections of the coloured regions. {A, B and C denote the flux domains corresponding to the flux tubes shown in Figure \ref{newfig}.} The green and purple dots denote where the separators cross this plane. (b) State 3a. Null graph and 3D field structure when an end null is bifurcated. (c) State 4. Null graph and 3D field structure when the central null is bifurcated. {Note that the connectivities shown in each null graph neglect the winding of different separators around one another.}}
\label{sep_diagram}
\end{figure*}

\subsection{Separator current layer breakup}
\subsubsection{Magnetic field connectivity}
Connecting the three null points along the top of the separatrix dome are a pair of separator field lines, {Fig. \ref{curtaintop}.} Like 3D null points, these are known to be preferred sites for current sheet formation and magnetic reconnection \citep[e.g.][]{longcope1996}. It has been previously observed that these sheets are prone to fragment, yielding a current layer containing multiple separators \citep{parnell2008,parnell2010a}. We thus begin by considering the effect of {super-imposing} one then more flux rings to simulate the effect of tearing in a reconnecting current layer around the separator. Such a model with a single flux ring was presented by \cite{wilmotsmith2011a} using a background field with two nulls, both with initially planar fan surfaces. They noted that as they increased the strength of the flux ring new separators appeared, coinciding with the formation of distinct new domains of magnetic flux connectivity. {By magnetic flux domain here and throughout we mean a volume within which there is a continuous change of field line connectivity. Distinct flux domains are bounded by separatrix surfaces at which this connectivity change is discontinuous.}

We observe the same effect when adding a single flux ring on the separator -- state 2a (see Table \ref{tbl}). Tracing field lines from the photosphere ($z=0$) and making a connectivity map as before, we observe the presence of a region of open flux nested within the closed field region, Figure \ref{sep_oc_1rope_phot}(a,b). Increasing the strength of the flux ring, we observe progressively more open and closed flux volumes being created, nested within one another -- Fig. \ref{sep_oc_1rope_phot}(c), state 2b. The formation of these new flux domains  corresponds to the formation of new pairs of separators joining the two associated nulls. Figure \ref{sep_diagram}(a) demonstrates this for state 2b. Whereas originally one separator joined the central and end nulls the formation of the three nested flux domains (Fig. \ref{sep_oc_1rope_phot}(b)) corresponds to the birth of three additional pairs of separators, giving seven in total (green and purple field lines), see below for a further discussion.



\begin{figure}[!t]
\centering
\includegraphics[width=0.45\textwidth]{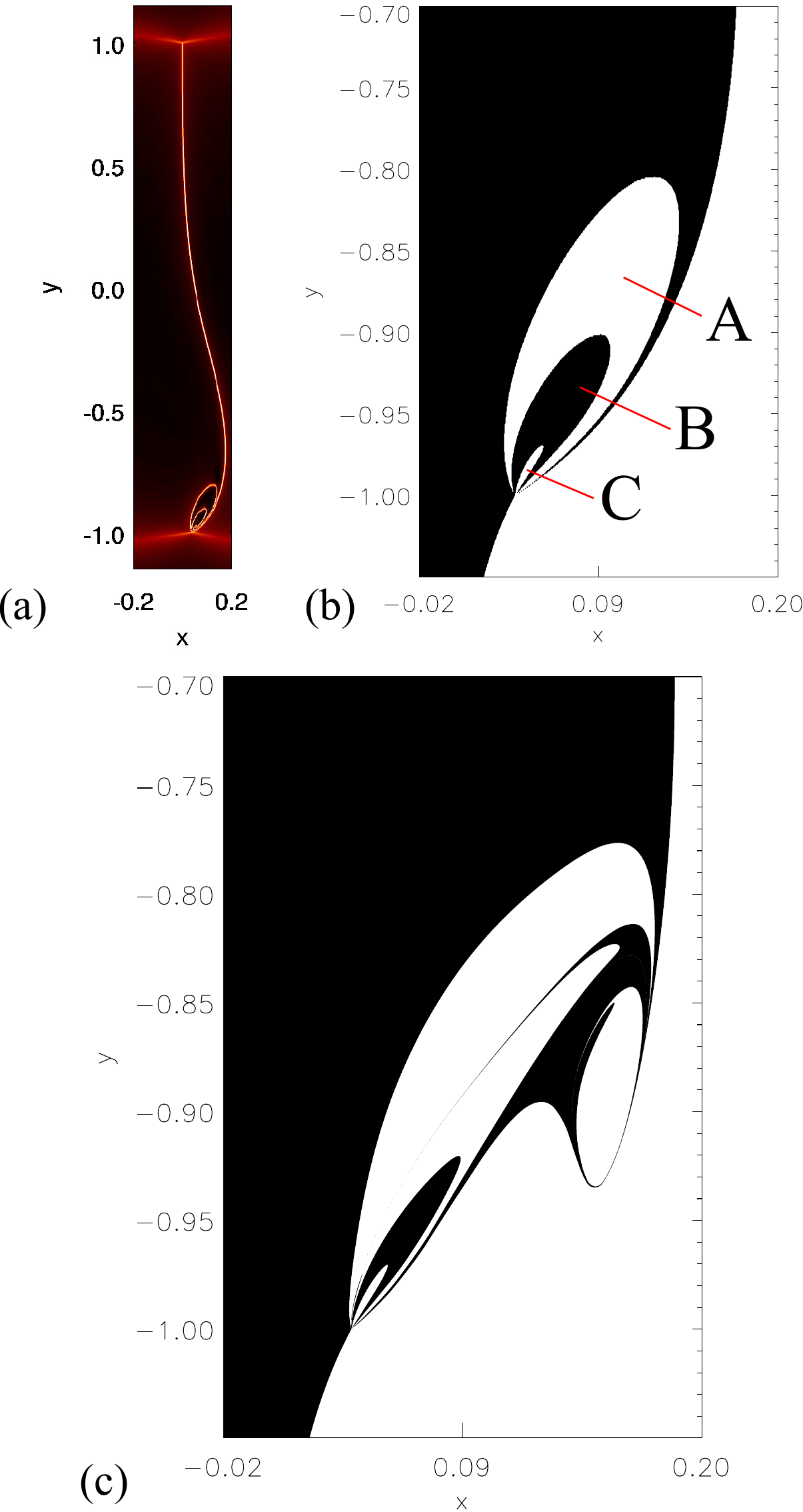}
\caption{(a) Squashing factor $Q$ on the surface $z=2$, and (b) close-up connectivity map, both for state 2b. The black region contains footpoints of field lines that connect to the photosphere at $x\leq 0$ (on one side of the separatrix dome/curtain structure), in the white regions field lines connect to $x>0$ (on the other side). The letters correspond to the letters marking the flux tubes in the 3D plot of Figure 10. (c) Connectivity map for state 2c.}
\label{sep_oc_1rope_top}
\end{figure}
We now turn to examine the connectivity of field lines that extend outwards into the heliosphere (those that exit through the top boundary). Throughout we consider the surface $z=2$ as being the `top' boundary -- field lines are close to vertical above this plane and so little deformation of the field line mapping occurs. In Figure \ref{sep_oc_1rope_top}(a) a map of $Q$ is plotted on the top boundary (as calculated between the two surfaces $z=0$ and $z=2$) for state 2b. Note that a colour scale is not shown since the maximum value is arbitrary, depending only on the resolution of the field line grid. We observe the imprint of the separatrix curtain, as well as additional nested loop structures that correspond to additional separatrix surfaces separating nested flux domains. In Figure \ref{sep_oc_1rope_top}(b) a connectivity map is plotted -- field lines that intersect the black region  connect to the photosphere at $x\leq 0$ on one side of the separatrix dome/curtain structure, while field lines intersecting the white region connect to the photosphere on the other side of the dome, $x>0$. {Embedding our structure in a global field these two different regions would correspond to {open field regions} of the same polarity that are disconnected at the photosphere \citep[see e.g.~Figure 5 of][]{platten2014}, and thus the figure shows that flux from the two disconnected coronal holes forms a mixed, nested pattern. These nested connectivity regions are entirely equivalent to those described in the photospheric connectivity maps above.} 
{It is expected that in a dynamic evolution there is continual reconnection of field lines within the current layer, and thus a mixing of plasma between all of the nested flux domains \citep[see e.g.][]{parnell2008}. As such, field lines at large height in these nested flux regions will continually be reconnected with those from the closed flux region.}

The addition of further flux rings -- representing further plasmoid structures in the reconnecting current layer -- leads to the formation of additional adjacent sets of nested open/closed flux domains. Figure \ref{sep_oc_1rope_top}(c) shows the connectivity map when three flux rings are present, state 2c. The complexity quickly becomes very high, with extremely thin layers of connectivity {with characteristic thickness of order $10^{-3}D$} -- even though the flux rope structures {and their collective footprint in the solar wind} {remain much larger, having diameters of order $10^{-1}D$. The inclusion of further flux ropes would decrease the length scales in the mapping yet further.}  This complexity of field line mapping was observed in a related context by \cite{parnell2010a}.

\subsubsection{Relation to 3D magnetic topology}\label{subsubtop}
{The direct association between the newly-created nested flux domains and additional separators that form in the domain is demonstrated in the right-hand frame of Figure \ref{sep_diagram}(a). Here we note that the seven separators lie at the intersections of the four different connectivity regions.}
The nested formation of flux regions and the associated pairs of separators may be understood as follows. 
Consider a bundle of field lines passing in along the open spine of the null at {$(0,-1,1)$}. As the flux ring strength is increased some of these field lines are wrapped repeatedly around the axis of the flux {rope before they reach the photosphere}. {Field lines reach the photosphere near spine {footpoints} of the central null, either at $x<0$ or $x>0$. They may do this directly, or by first winding once, twice, or more times around the flux rope axis. {This is demonstrated in Figure \ref{newfig}, where flux tubes are plotted from each of the nested connectivity domains that intersect the upper boundary of state 2b (marked `A', `B' and `C' in Figure \ref{sep_oc_1rope_top}(b))}. Each additional winding corresponds to a new flux domain.
 {This is because as the strength of the flux ring is increased the separatrix surfaces of the two nulls (red and cyan curves in the right panel of Figure \ref{sep_diagram}(a)) fold over to intersect with one another multiple times. Each additional pair of intersections correspond to a pair of new flux domains (bounded by portions of the separatrices) and a pair of new separators \citep[see the right-hand image of Figure \ref{sep_diagram}(a) and Figures 3 and 4 of][]{wilmotsmith2011a}.}
Field lines wind more times closer to the flux ring axis, and so new topological domains are} formed within the previous ones along with a pair of separators. This explains the nested nature of the connectivity domains observed in Fig.~\ref{sep_oc_1rope_phot} and why each new pair of separators have one half twist more than the {preceding} two, Fig.~\ref{sep_diagram}(a). 
\begin{figure}[!t]
\centering
\includegraphics[width=0.5\textwidth]{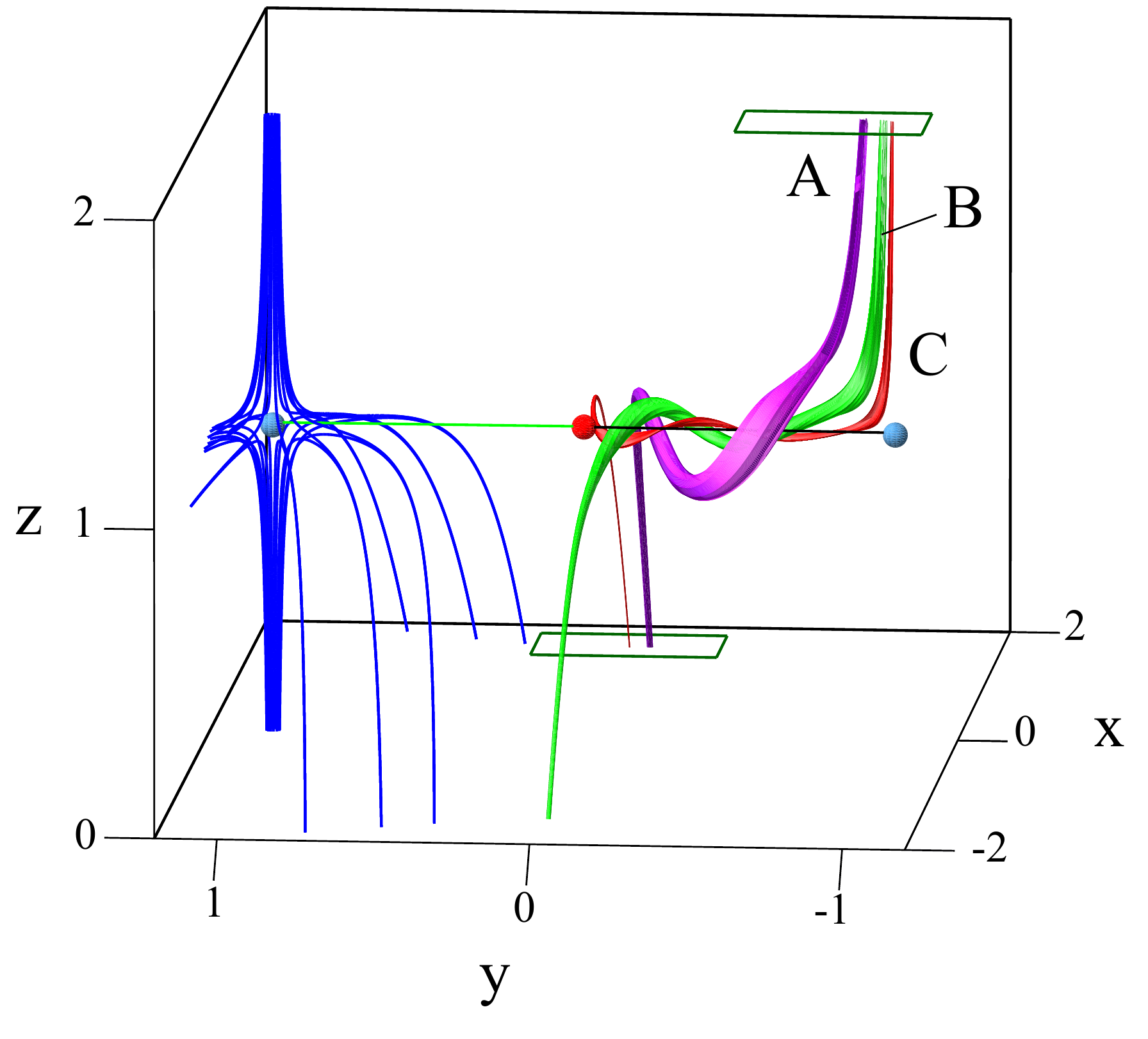}
\caption{{Flux tubes (red, green, magenta) located in the nested flux domains that intersect the upper boundary of state 2b. The letters correspond to the letters marking the flux domains in Figure \ref{sep_oc_1rope_top}(b).}}
\label{newfig}
\end{figure}

\subsection{Effect of null point bifurcations}
So far we modelled the case in which the current layer forms along the separator, with this current layer breaking up but the number of nulls remaining fixed -- that is, no null bifurcations occurred. However, another distinct possibility is that the current layer that forms in response to a dynamic driving of the system contains one or more of the coronal nulls. When such a current layer breaks up we would expect a bifurcation of the corresponding null point(s), which naturally should also coincide with the formation of additional separators -- indeed this may well occur even when the current is focussed away from the nulls, as observed by \cite{parnell2010a}.
\begin{figure*}[!t]
\centering
\includegraphics[width=\textwidth]{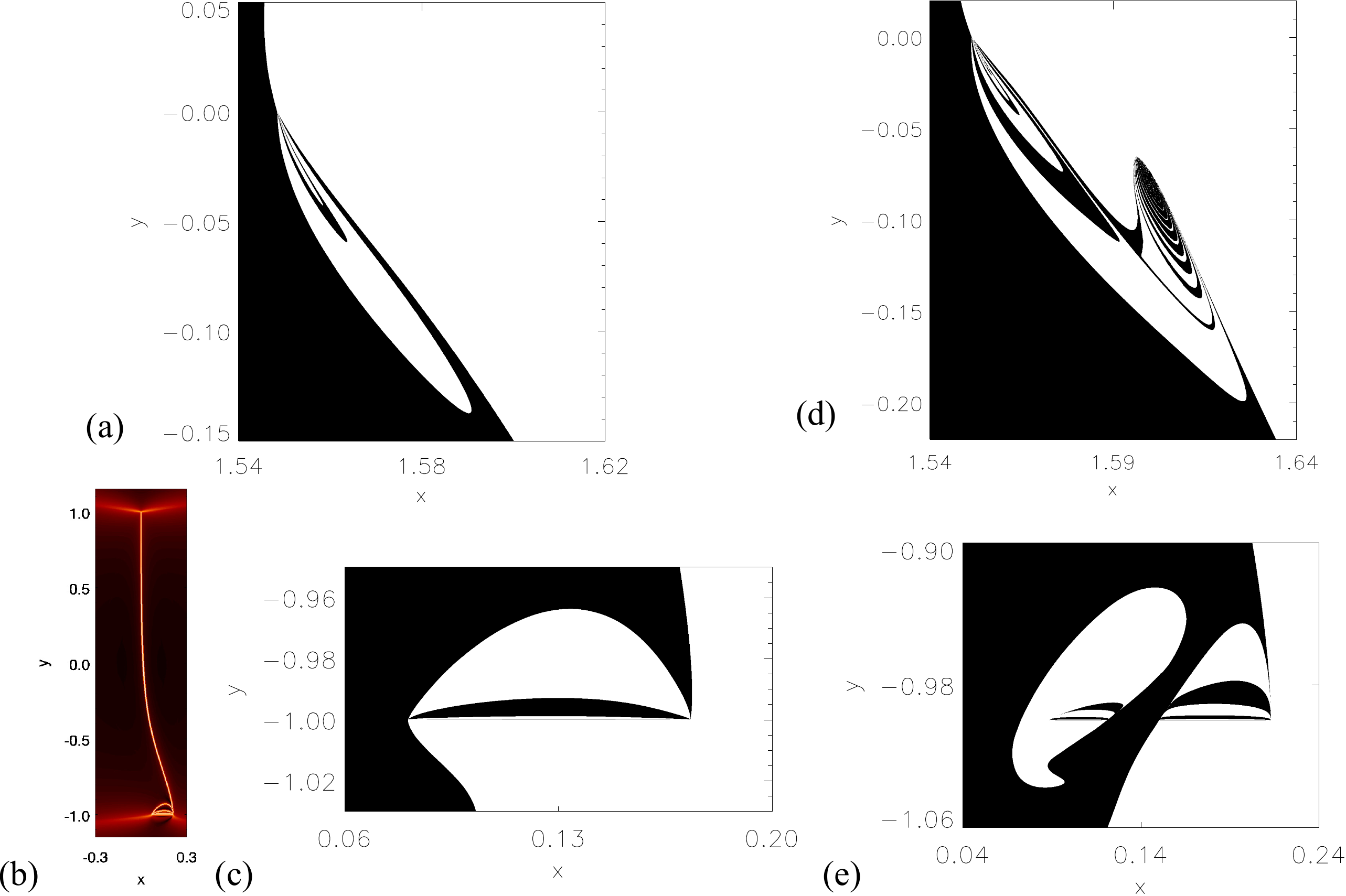}
\caption{{(a) {State 3a:} Close-up connectivity map on the photosphere with colours as in Figure \ref{sep_oc_1rope_phot}. (b) State 3a: $Q$-map on the top boundary. (c) {State 3a:} Close-up connectivity map on the upper boundary with colours as in Figure \ref{sep_oc_1rope_top}; (d) {Same as (a) for state 3b. (e) Same as (c) for state 3b.}}}
\label{sep_oc_1ropeEN}
\end{figure*}

We consider here two distinct cases, in the first of which the null point initially at $(x,y,z)\approx(0,-1,1,)$ is bifurcated into multiple null points, and in the second of which the central null (initially at (0,0,1)) is bifurcated. Consider first the situation where the end null is bifurcated. First, adding a single flux ring of sufficient strength at the initial location of the null we obtain a bifurcation to form three nulls as in Figure \ref{WPtop} (state 3a) {-- see Figure \ref{sep_diagram}(b)}. Let us now examine the result for the {magnetic flux connectivity, considering first flux intersecting the photosphere}. In Figure \ref{sep_oc_1ropeEN}(a) we see that the photospheric connectivity map appears as before: new nested open and closed flux domains are created in the vicinity of the spine footpoints of the central null. The connectivity map for {open flux traced from} the upper boundary  {is shown in Figure \ref{sep_oc_1ropeEN}(c) (where the colours have the same meaning as before). As shown in the $Q$-map (Figure~\ref{sep_oc_1ropeEN}b)}, the main separatrix curtain is diverted in the positive $x$-direction for negative $y$. It terminates on the spine of one of the null points located in the vicinity of $(x,y,z)=(0,-1,1)$. There is then an additional separatrix footprint orthogonal to this bounded by the spines of the newly created nulls as in Figures \ref{WPtop}, \ref{dome_q_map}. Interestingly though, the arcs of high $Q$ emanating from this separatrix (as in Figure \ref{dome_q_map}) now form the boundaries of the flux domains that connect to opposite sides of the dome footprint. 
{This can be understood by considering that {each arc} of high $Q$ in Fig.~\ref{dome_q_map} represents a further half turn of field lines along the axis of the flux rope, i.e.~the {outermost two} arcs correspond to field lines exiting along one or other spine of the central null  having wound {up to} once around the flux rope axis, the next pair to field lines that first wind {between once and} twice around the flux rope axis, and so on}. When these field lines are mapped on to the photosphere as in the domed single null case this leads to the continuous but rapid change in connectivity denoted by the $Q$ ridges. When such field lines separate along the spine of a distant null the change in connectivity becomes discontinuous, forming the nested flux domains, {see also Section \ref{subsubtop}.}


As before, the connectivity maps quickly become significantly more complex when additional flux ropes are added. Figures  \ref{sep_oc_1ropeEN}(d,e) show the photospheric and upper boundary connectivity maps when an additional two flux rings are added to generate a bifurcation to a state with three flux ropes pairs and seven nulls, state 3b. {Again, characteristic length scales of the mapping layers of order $10^{-3}D$ or below are observed within a mixed flux region with dimensions of order $10^{-1}D$.}

\begin{figure}[!t]
\centering
(a)\includegraphics[height=5cm]{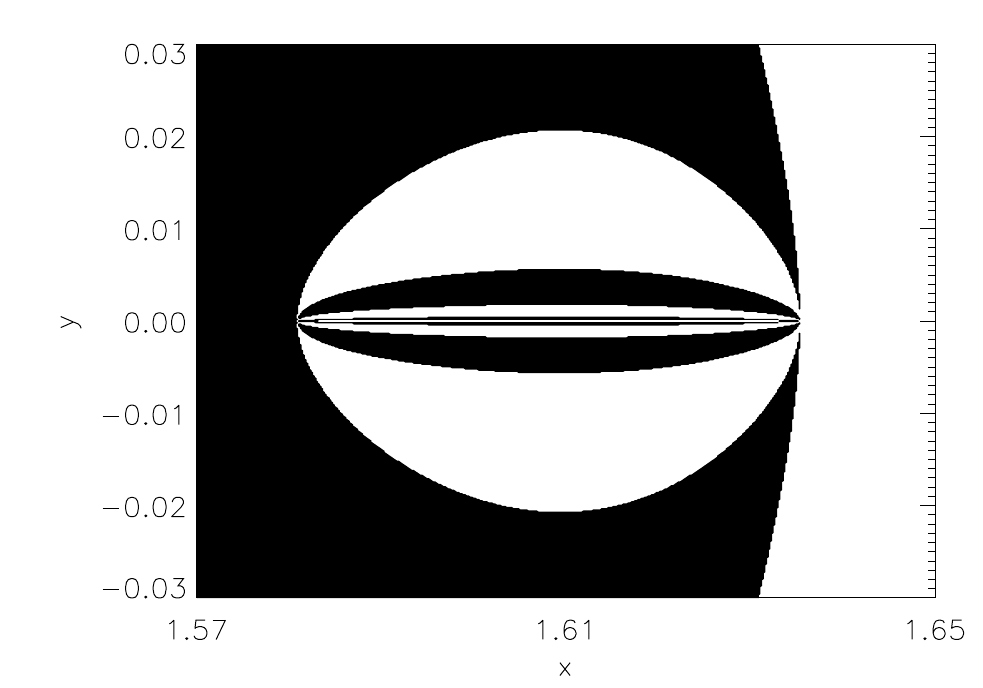}\\
(b)\includegraphics[height=5.4cm]{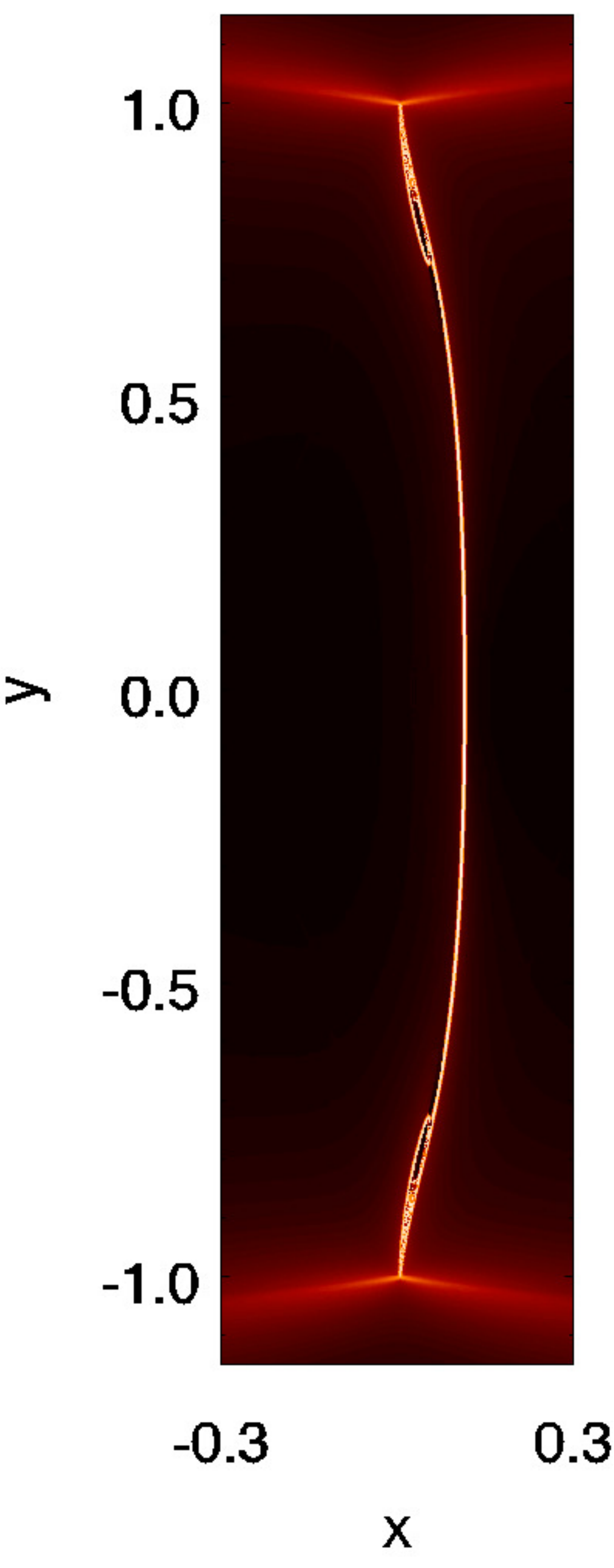}
(c)\includegraphics[height=5.4cm]{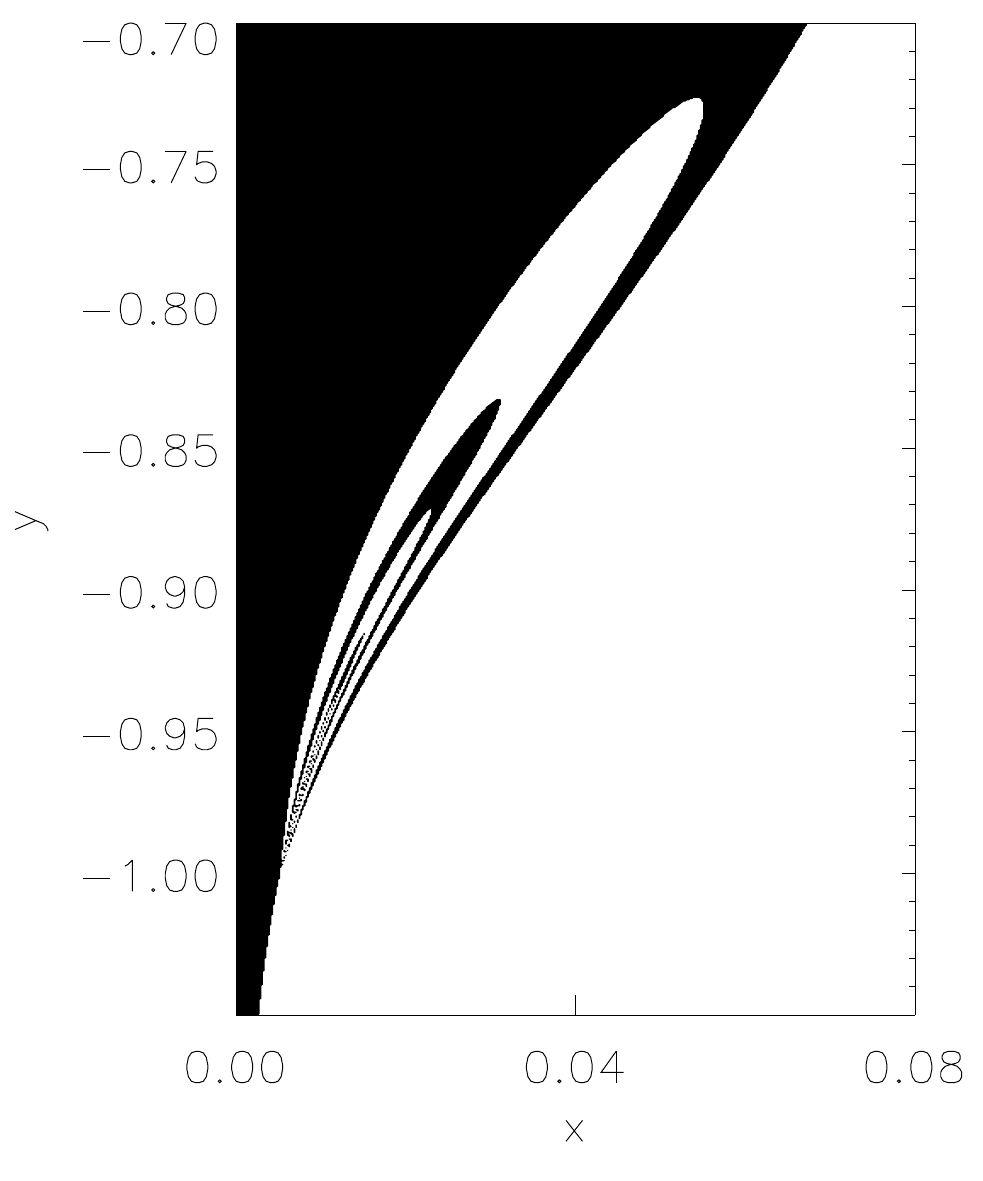}
\caption{All for state 4: (a) Close-up connectivity map on the photosphere with colours as in Figure \ref{sep_oc_1rope_phot}. (b) $Q$ on the upper boundary. (c) Close-up connectivity map on the upper boundary with colours as in Figure \ref{sep_oc_1rope_top} -- note that the $x$-direction is stretched in (c) for clarity.}
\label{sep_oc_1ropeCN}
\end{figure}
Finally, suppose that the fragmentation of the coronal current layer leads to a bifurcation of the central null point. Adding a single flux ring leads to a bifurcation to a state with three null points (state 4) as before. Analysis of the resulting topology reveals a situation that mirrors state 3a. As shown in Figure \ref{sep_oc_1ropeCN}(a), this time the photospheric connectivity maps show adjacent crescent-shaped domains of open and closed flux, symmetric about $y=1$ since the null bifurcation now leads to a bifurcation of both of the initial separators. Correspondingly, nested flux domains of alternating connectivity are now observed in the connectivity map for the upper boundary (Figure \ref{sep_oc_1ropeCN}(b)), this time emanating from the footpoints of both of the vertical open spines.

\section{Discussion}\label{discusssec}
\subsection{Mixing of open and closed flux}
As shown in the above models, reconnection at the Sun's open-closed flux boundary can result in that boundary taking on a highly non-trivial structure. In the presence of an isolated null point {separatrix dome no new flux domains are created but} an envelope forms around the initial dome structure in which magnetic flux from inside and outside the dome is efficiently mixed together {in spiral patterns}. As shown by \cite{wyper2014a} magnetic flux is continually and recursively reconnected from open to closed and back again within this envelope 
{\citep[i.e.~is reconnected back and forth multiple times between open and closed regions --][]{parnell2008}}. 
The result for the field at large heights is that a flux tube is present around the original spine line within which field lines are being continually reconnected with those from the closed region beneath the dome. 

If we consider a more complicated structure in which coronal separators are present, the breakup of the current layer leads to the formation of new flux domains. In particular, {new open and closed magnetic flux domains form in nested structures}, whose  length scales become rapidly {shorter; even for the models considered here containing just three {flux rope pairs} characteristic length scales of the mapping layers of order $10^{-3}D$ or smaller are observed}. The expectation is that in a dynamic evolution, continual {transfer of flux}/plasma between the narrow open and closed layers would occur. The new regions of flux are observed to form in the vicinity of the footpoints of spine field lines in the pre-reconnection field. {Together they cover a region of comparable scale to the distribution of current and flux rope structures, here of order $10^{-1}D$}.

\subsection{Implications for solar wind models}
Our results imply that in the vicinity of open spine structures and open separatrix curtain structures, an efficient mixing of open and closed magnetic flux, and the associated plasma, is likely to take place whenever reconnection occurs at the corresponding nulls or separatrices. This is an attractive ingredient for explaining observed properties of the slow solar wind by the interchange reconnection model. In particular, the slow solar wind is known to be highly fluctuating in both composition and velocity (in both space and time), with the composition properties varying from close to those of the closed corona to nearly photospheric \citep{geiss1995,zurbuchen2007}. Contributing factors to this fluctuating, filamentary structure} could be the bursty nature of the interchange reconnection, {and the complex spatial structuring on large and small scales of the open-closed boundary. Combining our results with those from simulations of current layer instabilities, it is clear that the reconnection process should lead} to a highly dynamic magnetic topology in which regions of open and closed flux are born and evolve in a complex pattern. 

{Interchange reconnection models for solar wind acceleration share the common feature that they require regions of open flux at photospheric heights that are at least predominantly surrounded by closed flux. This is consistent with observations of significant components of solar wind outflow emanating from locations adjacent to active regions \citep[e.g.][]{neugebauer2002,woo2004,brooks2015}. The models of Fisk and co-workers take a statistical approach to the evolution of the Sun's open flux, in which they assume a random orientation for the closed loops and thus an isotropic diffusion of open field lines as they random walk through the closed flux region. One piece of observational evidence that is argued to support their model of open field dynamics is the coincidence of extended open field regions with minima in the local rate of flux emergence \citep{fisk2005,abramenko2006,hagenaar2008}.
Interestingly, the models also predict that the random walk of open field lines will induce a braiding of field lines in the heliosphere, and this is also a prediction of the MHD simulations of \cite{wyper2014a,wyper2014b} (see Figure \ref{sim_q_map}), though in our model the braiding is induced by turbulent dynamics in the reconnecting current layer.

The S-web model, by contrast to the models of Fisk and co-workers, seeks to identify explicitly the locations of possible interchange reconnection -- and thus outflow -- by analysing the detailed magnetic field topology. In a given magnetic field extrapolation open field channels and patches can be identified, however there are indications that the number of disconnected open field regions can greatly increase when the resolution of the photospheric magnetogram is increased \citep{edwards2014}. Such an increase corresponds to an increase in the number of arcs in the S-web. In this paper, we have argued that when interchange reconnection occurs at a structure of the open-closed boundary -- and the full reconnection dynamics are included -- it will tend to generate a layer within which open and closed flux are mixed. So in this picture, each separatrix arc of the S-web becomes not a single line but a band within which the open-closed flux boundary is highly structured. This may partially mitigate against the {concern} that the static S-web is not space-filling and therefore cannot provide a continuous slow solar wind outflow \citep{wang2012}. 
 The dimensions of this layer of mixed open-closed flux} are of course crucial, but cannot be readily estimated from the present approach. They will depend both on the size and geometry of the fragmented reconnecting current layer, and the overall global field geometry that connects this volume out to the heliosphere in any given situation. While the latter can be estimated from static models, elucidating the former will require a full, detailed dynamical understanding of the reconnection process. 
 {We note also that our results perhaps provide a `bridge' between the Fisk et al.~and S-web models, in the sense that the broad bands of {mixed open-closed flux} expected to form around the arcs of the S-web {can} be thought of as a regions within which the diffusive random walk envisaged by Fisk and co-workers becomes highly efficient. The indication is then that the broadly uniform diffusion coefficient in those models should in fact be highly structured in both space and time. However, to determine the spatial and temporal distribution of these regions of efficient open-closed flux mixing {-- and thus their overall effect --} will require a statistical study of the structures of the open-closed flux boundary, together with the results of the dynamical simulations suggested above. }


\subsection{Morphology and structure of flare ribbons}
We now consider the implications of our results for signatures of particle acceleration in topologies involving coronal null points and separators. Of course, explaining SEP and flare ribbon observations requires knowledge of particle behaviour, and that remains to be studied. However, the above results allow us to make a number of predictions. Consider first impulsive SEP events. A series of recent observational studies has shown that impulsive SEP sources are located in open field regions magnetically well-connected with the target \citep[][and references therein]{reames2013}. However, since their composition is more indicative of closed field regions, it has been proposed that they are accelerated directly during interchange reconnection in the low corona \citep{drake2009}. The present results demonstrate that particles accelerated (by some means) during the reconnection process readily have access to open field lines, since all field lines are recursively reconnected from open to closed within the envelope of mixed flux described above. What's more, when open and closed flux are mixed into sufficiently thin layers, the distinction between the two is lost for the particles -- {certainly the case if the layers are thinner than the Larmor radius. Furthermore, the dynamics in the current layer are most likely to be turbulent, and braiding of field lines is expected \citep{wyper2014b}, both of which are known to lead to enhancement of cross-field particle transport \citep[e.g.][and references therein]{kontar2011}.}

Our results, combined with those of \cite{wyper2014a}, also provide insights into the expected structure of flare ribbons in coronal null point and pseudo-streamer topologies. In particular, it has been shown that the separatrix and QSL footprints often map onto the locations of the flare ribbons. However, the flare ribbons usually exhibit additional structure, often bright kernel-like structures that move along the ribbons \citep{nishizuka2009,barta2011}. These features could correspond to the footpoints of the flux rope structures formed during 3D current sheet fragmentation, these being associated with bundles of efficiently mixed open and closed flux. If this were the case, one would expect the motion of the bright features to be linked to the velocity of the outflow from the reconnection region, multiplied by some factor resulting from the geometry of the magnetic connection between the reconnection site and the photosphere. However, as shown by \cite{wyper2014b}, the flux ropes exhibit a complicated dynamics as they kink and interact with one another in the reconnection region, so the motion of their footprint on the photosphere is expected to deviate significantly from a simple advection. {We note that performing the same procedure as herein -- adding a flux ring to simulate the effect of reconnection -- for a background field defined by a QSL or hyperbolic flux tube leads not to more structure in the associated $Q$-map, but simply to a break of the high-$Q$ layer (results not presented here). This may correspondingly imply that the signatures of bursty reconnection in such a field topology are different to the case when separatrices are present.}

One further specific conclusion that can be drawn regards the nature of the  flare ribbons associated with the spine footpoints in the isolated dome topology. \cite{masson2009} and \cite{reid2012} reported observations of the flare ribbons in such a topology and noted that the ribbons that were postulated to be connected with the spine footpoints were extended structures with high aspect ratio {(rather than being circular)}. They proposed that this was related to the distribution of the squashing factor for field lines around the spine in the associated extrapolated equilibrium field. The results of Section \ref{domesec} provide us with an alternative hypothesis: the extended elliptical ribbons may mark out the imprint of the vertical separatrix curtains (and surrounding arcs of high $Q$) associated with a fragmented current layer. It may also be that both effects are important for the spine footpoint ribbon extension. Of course it remains to be seen what accelerated particle distributions are expected during a dynamic reconnection event, and this will be pursued in a future study.

\section{Summary and conclusions}\label{concsec}
Magnetic reconnection in the solar corona is likely to occur in highly fragmented current layers, {as demonstrated in recent 3D simulations \citep{daughton2011,wyper2014a}. Here we have used simple static magnetic field models to investigate the implications of the current layer fragmentation on the large-scale topology of representative solar coronal field structures.} We have shown that this fragmentation can vastly increase the topological complexity beyond that of the equilibrium magnetic field. In particular, when the fragmenting current layer forms at the open-closed magnetic flux boundary, the structure of that boundary can become highly complex. The results, however, are also relevant for the studied topologies in the case where all flux is globally closed -- in this case some flux closes at some distant point on the photosphere (but is `locally open').

{We considered here two principal topologies; the isolated null point dome and the separatrix curtain topology. Both of these are observed over a broad range of characteristic scales in the corona, from hundreds of Mm down to tens of km.}
In the presence of an isolated dome, non-linear tearing of the reconnecting current layer leads to the formation of an envelope of magnetic flux around the initial dome structure in which flux from inside and outside the dome is efficiently mixed together. Magnetic flux is continually recursively reconnected from open to closed and back again within this envelope. The result for the field at large heights is that a flux tube is present around the original spine line within which field lines are being continually reconnected with those from the closed region beneath the dome. Such isolated dome structures are typically found in abundance in coronal holes in magnetic field extrapolations.
In the separatrix curtain (`pseudo-streamer') topology of Section \ref{curtainsec} we saw that the breakup of the current layer leads to the formation of new flux domains. In particular, open and closed flux {(as well as flux from pairs of disconnected open field regions)} form in nested domains with very short length scales. {The thickness of the adjacent open and closed flux domains can be many orders of magnitude smaller than the global length scale of the field structure or indeed of the flux ropes in the current layer -- in our models with only three flux ropes the mapping layers were two orders of magnitude smaller than the flux ropes.} The expectation is that in a dynamic evolution, continual reconnection between the narrow layers of open and closed flux would occur {within a flux envelope surrounding the new nested flux domains. Our static models predict that in the corona immediately above the pseudo-streamer this envelope will cover a region of comparable scale to the distribution of current and flux rope structures. However, this would be expected to widen with height as the field strength reduces with radial distance from the sun and the field expands laterally}.

Understanding particle acceleration in topologies such as those studied could help us comprehend both the  release of impulsive SEPs to open field lines and the appearance of certain flare ribbons. In particular, the appearance of extended flare ribbons at spine footpoints in the dome topology could be related to the separatrix footprints that appear there when null point bifurcations occur as the current layer fragments. What's more, the efficient mixing of open and closed flux in the reconnection process provides a natural mechanism for accelerated particles to access the open field region. Future studies of particle acceleration during the reconnection process will reveal much more.

Global magnetic field extrapolations are now revealing the huge complexity of the coronal  field, and in particular the structure of the boundary between open and closed magnetic flux. Regions of open flux that are either disconnected from the polar coronal holes at the photosphere or connected only by narrow open-flux corridors contribute arcs to the S-web \citep{antiochos2011}. Our results show that whenever reconnection occurs at a null point or separator of the open-closed boundary, the associated separatrix arc of the S-web becomes not a single line but a band of finite thickness within which the open-closed flux boundary is highly structured. The dimensions of this band are of course crucial, but cannot be readily estimated from the present approach. The next step then, to determine the importance of this effect, requires dynamical MHD simulations of the process in order to quantify the dimensions of this band and the flux associated with it.

\acknowledgments
DP acknowledges financial support from the UK's STFC (grant number ST/K000993) and the Leverhulme Trust. PW acknowledges support from an appointment to the NASA Postdoctoral Program at Goddard Space Flight Center, administered by Oak Ridge Associated Universities through a contract with NASA.


\clearpage

\begin{table}
\begin{center}
\caption{Details of model field configurations. $s_n$ is the strength of the $n^{th}$ flux ring ($B_0$ in Equation \ref{b_ring}), $p_n$ is the position of its centre ($(x_N,y_N,z_N)$ in Equation \ref{b_ring}).\label{tbl}}
\begin{tabular}{cccccccccccc}
\tableline\tableline
State & $\BB$-field\tablenotemark{a}  & $s_1$ & $p_1$ & ${{L_{1}}^2}$ & ${{l_{1}}^2}$ & $s_{2,3}$ & $p_2$  & $p_3$ & ${{L_{2,3}}^2}$ & ${{l_{2,3}}^2}$ \\
\tableline
1a & dome  & 0.03  &$\left(\begin{array}{c}0\\0\\0.5\end{array}\right)$ & 0.025 & 0.02 & -- & -- & -- & -- & --   \\
1b & dome  & 0.05  &$\left(\begin{array}{c}0\\0\\0.5\end{array}\right)$ & 0.025 & 0.02 & -- & -- & -- & -- & --   \\
1c & dome  & 0.08 & $\left(\begin{array}{c}0\\0\\0.5\end{array}\right)$ & 0.025 & 0.02& 0.05 & $\left(\begin{array}{c}0.158\\0\\0.380\end{array}\right)$  & $\left(\begin{array}{c}-0.158\\0\\0.604\end{array}\right)$ & 0.005 & 0.005 \\
2a & curtain  & 0.18 & $\left(\begin{array}{c}0\\-0.5\\1.0\end{array}\right)$ & 0.1 & 0.167 & -- & -- & -- & -- & --  \\
2b & curtain  & 0.25 & $\left(\begin{array}{c}0\\-0.5\\1.0\end{array}\right)$ & 0.1 & 0.167 & -- & -- & -- & -- & --  \\
2c & curtain  & 0.25 & $\left(\begin{array}{c}0\\-0.5\\1.0\end{array}\right)$ & 0.1 & 0.167 & 0.06 & $\left(\begin{array}{c}-0.170\\-0.500\\0.822\end{array}\right)$ & $\left(\begin{array}{c}0.171\\-0.500\\1.142\end{array}\right)$ & 0.005 & 0.167  \\
3a & curtain  & {0.22} & $\left(\begin{array}{c}0\\-1.0\\1.0\end{array}\right)$ & 0.1 & 0.333 & -- & -- & -- & -- & --  \\
3b & curtain  & 0.3 & $\left(\begin{array}{c}0\\-1.0\\1.0\end{array}\right)$ & 0.1 & 0.333 & 0.05 & $\left(\begin{array}{c}-0.145\\-1.000\\0.832\end{array}\right)$ & $\left(\begin{array}{c}0.132\\-1.000\\1.131\end{array}\right)$ & 0.0025 & 0.333  \\
4 & curtain  & 0.2 & $\left(\begin{array}{c}0\\0\\1.0\end{array}\right)$ & 0.1 & 0.333 & -- & -- & -- & -- & --  \\
\tableline
\end{tabular}
\tablenotetext{a}{Background magnetic field where ``dome" corresponds to Equation (\ref{bdome}) and ``curtain" to Equation (\ref{bcurtain}).}
\end{center}
\end{table}

\end{document}